\documentclass[12pt]{article}
\pdfoutput=1

\usepackage[utf8]{inputenc}
\usepackage[left=2.55cm, right=2.55cm, top=2.55cm, bottom=2.55cm]{geometry}
\usepackage{amsmath,amssymb,amsbsy,mathtools}
\usepackage{slashed}
\usepackage{xcolor}
\usepackage{graphicx}
\usepackage{url}
\usepackage{cancel}
\usepackage{cite}
\usepackage[colorlinks=true,allcolors=blue,pdfborder={0 0 0},linktocpage=false,pdfencoding=auto]{hyperref}
\usepackage{tabularx,booktabs}
\usepackage{multicol}
\usepackage{feynmp}
\usepackage{units}
\usepackage{xspace}
\usepackage[labelfont=bf]{caption}
\usepackage[section]{placeins}
\usepackage{subcaption}
\usepackage{soul}
\usepackage{bbold}
\usepackage{parskip}
\usepackage{float}
\usepackage{tabulary}
\usepackage{empheq}
\usepackage{tabu}

\definecolor{darkred}{rgb}{0.6,0,0}
\definecolor{darkpurple}{rgb}{0.5,0,0.5}

 \newcommand{\code}[1]{\texttt{#1}}
\newcommand{\beqn}{\begin{eqnarray}}
\newcommand{\eeqn}{\end{eqnarray}}

\begin{document}
\author{Amin Aboubrahim$^{a}$\footnote{\href{mailto:abouibra@union.edu}{abouibra@union.edu}}~~and Pran Nath$^b$\footnote{\href{mailto:p.nath@northeastern.edu}{p.nath@northeastern.edu}}
 \\~\\
$^{a}$\textit{\normalsize Department of Physics and Astronomy, Union College,} \\
\textit{\normalsize 807 Union Street, Schenectady, NY 12308, U.S.A.} \\
$^{b}$\textit{\normalsize Department of Physics, Northeastern University,} \\
\textit{\normalsize 111 Forsyth Street, Boston, MA 02115-5000, U.S.A.} \\
}

\title{\vspace{-2cm}
\vspace{1cm}
\large \bf Upper limits on dark energy-dark matter interaction from DESI DR2 in a field-theoretic analysis
 \vspace{0.5cm}}
\date{}
\maketitle

\vspace{0.5cm}

\begin{abstract}
One of the important issues both in particle physics and cosmology relates to whether dark energy is a cosmological constant $\Lambda$, or is  dynamical in nature such as quintessence. In this work, we discuss a model of quintessence interacting with dark matter and analyze the resulting phenomenology of the dark energy equation of state. We identify two regions where the equation of state behaves differently depending on the size of the dark matter-dark energy interaction strength. We show that the strong coupling region induces a transmutation of quintessence from thawing to freezing. Using the recent data release from the Dark Energy Spectroscopic Instrument (DESI), we rule out this possibility of transmutation and investigate the weak coupling region to derive upper limits on the interaction strength. Our analysis indicates that while $\Lambda$CDM lies within the $1\sigma$ contour in the $w_0$-$w_a$ plane, the best fit points lie in the fourth quadrant and show deviations from the $\Lambda$CDM prediction. 

\end{abstract}

\numberwithin{equation}{section}

\newpage

{  \hrule height 0.4mm \hypersetup{linktocpage=true} \tableofcontents
\vspace{0.5cm}
 \hrule height 0.4mm}

\section{Introduction}

Currently the concordance $\Lambda$CDM model is considered the Standard Model of cosmology and is successful at explaining a large amount
of cosmological data. It is described by the action 
\begin{align}
S_{\Lambda \rm CDM}=  \int \text{d}^4x \sqrt{-g} 
 \left[\frac{1}{16\pi G}(R-2 \Lambda)+ {\cal L}_{\rm CDM}\right], 
 \label{eq1}
\end{align}
where $g=\text{det}(g^{\mu\nu})$ is the determinant of the metric $g^{\mu\nu}$. We use the Friedmann-Lema\^itre-Robertson-Walker metric in conformal time $\tau$ so that the line element is given by 
$\text{d}s^2=a^2(\tau)\left[-\text{d}\tau^2+g_{ij}\text{d}x^i \text{d}x^j\right]$, where $a(\tau)$
is the scale factor. In Eq.~(\ref{eq1}), $R$ is the Ricci scalar, $G$ is Newton's constant, $\Lambda$ is the Einstein cosmological constant, and ${\cal L}_{\rm CDM}$ is the Lagrangian density for Cold Dark Matter (CDM). More generally, ${\cal L}_{\rm CDM}$ can be replaced by the Lagrangian density ${\cal L}_{\rm m}$ which includes all forms of matter and radiation. In spite of the great success of Eq.~(\ref{eq1}), some anomalies have recently emerged, with
two of the prominent ones being the Hubble tension, $H_0$, and the degree of matter clustering, $S_8$~\cite{Abdalla:2022yfr,DiValentino:2025sru}.  
Several suggestions to reduce the Hubble tension have been proposed which, to name a few, include interacting dark matter models~\cite{Pourtsidou:2013nha,Pourtsidou:2016ico,Linton:2017ged,Chamings:2019kcl,Pan:2019gop,Bonici:2018qli,Yang:2019uzo,Pan:2020zza,Wang:2016lxa,Wang:2024vmw,Boehmer:2015kta,Boehmer:2015sha,Potter:2011nv,DiValentino:2019jae,DiValentino:2019ffd,Escamilla:2023shf,Bernui:2023byc,Yang:2022csz,Amendola:1999er,Kase:2019veo,Perez:2021cvg,Garcia-Arroyo:2024tqq,Beyer:2010mt,Beyer:2014uqa,vandeBruck:2022xbk,Yang:2018euj,Lee:2022cyh}, early dark energy~\cite{Doran:2006kp,Agrawal:2019lmo,Vagnozzi:2023nrq}, decaying dark matter~\cite{Berezhiani:2015yta,Ibarra:2013cra} as well as models introducing extra relativistic degrees of freedom~\cite{Aboubrahim:2022gjb,Fernandez-Martinez:2021ypo,Escudero:2021rfi,Baumann:2015rya,Brust:2017nmv,Blinov:2020hmc,Jacques:2013xr,Cuesta:2021kca,Vagnozzi:2019ezj}.
A current trend is to introduce an interaction term (or a source term) between dark matter  (DM) and dark energy (DE) at the level of the continuity equations~\cite{Abdalla:2022yfr,Valiviita:2008iv,Gavela:2009cy,DiValentino:2019jae}, so that
\begin{align}
\label{eq2}
&\rho^\prime_\chi+3\mathcal{H}(1+w_\chi)\rho_\chi=Q,\\
&\rho^\prime_\phi+3\mathcal{H}(1+w_\phi)\rho_\phi=-Q.
\label{eq3}
\end{align}
Here $\rho_\chi$ and $\rho_\phi$ are the energy densities for 
DM and DE, $w_\chi$ and $w_\phi$ 
are the corresponding equations of state (EoS), with $w$ defined as the ratio of pressure $(p)$ to energy density
($\rho$) for a given species, $w=p/\rho$, and $^\prime\equiv \text{d}/\text{d}\tau$. However, in Eqs.~(\ref{eq2}) and~(\ref{eq3}), the energy conservation is imposed in an ad hoc manner by choosing the sources to be $+Q$ and $-Q$ and these equations are inconsistent within a field theoretic framework.

Other than the $H_0$ and $S_8$ tensions, other anomalies have emerged as well, specifically one related to the dark energy EoS. Thus, the Dark Energy Spectroscopic Instrument (DESI)~\cite{DESI:2024mwx,DESI:2025zgx} recent results have pointed toward a time-varying EoS. In $\Lambda$CDM, the cosmological constant's EoS is $w_\Lambda=-1$ and so a time-varying $w$ may indicate some underlying new physics deviating from $\Lambda$CDM. Using the Chevallier-Polarski-Linder (CPL) parameterization~\cite{Chevallier:2000qy,Linder:2002et} of the EoS, $w(a)=w_0+(1-a)w_a$, DESI DR1 reports a preference for $w_0>-1$ and $w_a<0$ with the $\mathrm{DESI~BAO}+\mathrm{CMB}+\mathrm{DESY}5$ data set at the $3.9\sigma$ level which grew to $4.2\sigma$ with DR2. Quintessence is a well known alternative to the cosmological constant which can provide a scenario for a dynamical EoS. Many models have been proposed in this direction leading to several classifications of quintessence models. Thus in thawing quintessence~\cite{Scherrer:2007pu,Caldwell:2005tm,Wolf:2023uno,Wolf:2024eph}, the field is initially frozen due to the Hubble friction so that $w_\phi\simeq -1$. The mass of the field then becomes smaller than $\mathcal{H}$ at late times causing a deviation of $w_\phi$ away from $-1$. In freezing models, $w_\phi$ decreases towards $-1$ at late times because of the shallow nature of the potential which, for a scaling freezing model~\cite{Ferreira:1997hj,Copeland:1997et}, is taken to be a double exponential potential. Compared to CPL, these models produce more complicated parameterizations of the EoS which do not lead to the crossing of the phantom divide, something the CPL suffers from.   

In this work, we examine the impact of a field-theoretic model on cosmology, and in particular, we focus on the dark energy equation of state. Motivated by a previous work~\cite{Aboubrahim:2024spa,Aboubrahim:2025qat}, where the authors have shown that such a field-theoretic approach can alleviate the $H_0$ tension and resolve the $S_8$ tension, we set out to explore the possibility of a dynamical EoS. Furthermore, we drop Eqs.~(\ref{eq2}) and~(\ref{eq3}) which are inconsistent 
within a field theoretic framework and consider a field theoretic formulation 
where energy conservation is automatic. We show that in this framework
DM-DE interaction can lead to a rich EoS phenomenology, where for a strong coupling, a transmutation of quintessence from thawing to scaling freezing occurs while a weaker coupling retains the thawing nature of quintessence. We investigate fits to the EoS in the two coupling regimes and examine the impact of the DM-DE coupling in bringing the model to agreement with DESI DR2. We show that while the DM-DE interaction can be an impactful parameter in aligning model predictions with DESI's results, a strong preference exists for non-interacting quintessence to explain DESI's DR2 results.

\section{The model}

In the analysis here we propose an alternative to $\Lambda$CDM of Eq.~(\ref{eq1}) which is a model of quintessence interacting with dark matter (QCDM) based on a Lagrangian. Thus we consider a set of  $n$ interacting spin zero fields 
$\phi_i$ ($i=1\cdots n$) where one of the fields is the quintessence field and the
rest are matter fields which may contain one or more dark matter fields and the 
rest may be other forms of matter fields. The action of this theory is given by
\begin{align}
 &S_{\rm QCDM}=\int \text{d}^4 x\sqrt{-g}
  \left[\frac{1}{16\pi G}R+ {\cal L}_{\rm QCDM}\right],
\end{align}
and the QCDM Lagrangian is
\begin{align}
 & {\cal L}_{\rm QCDM}=  \sum_i\frac{1}{2}\phi_i^{,\mu}\phi_{i,\mu}-
  V(\{\phi_i\},\{\phi_j\}),
\end{align}
where
\begin{align}
  &V=\sum_i^n V_i(\phi_i)+ V_{\rm int}(\{\phi_i\},\{\phi_j\}), \\
 &V_{\rm int}(\{\phi_i\},\{\phi_j\}) =\frac{1}{2} \sum_{i\neq j}\sum_jV_{ij}(\phi_i,\phi_j),
\label{n2}    
\end{align}
with $V_{ij}(\phi_i,\phi_j)$ being the interaction potential between the fields $\phi_i$ and $\phi_j$. The Klein-Gordon equation for the field $\phi_i$ is given by
\begin{equation}
\phi_i^{\prime\prime}+2\mathcal{H}\phi_i^\prime+a^2\left(V_{i,\phi_i}+
  \frac{1}{2} \sum_{j\neq i} (V_{ij}+ V_{ji})_{,\phi_i}\right)=0,
    \label{n3}
\end{equation}
where $V_{ij,\phi_i}\equiv \partial_{\phi_i}V_{ij}$, and  $\mathcal{H}=a^\prime/a$ is the conformal Hubble parameter. Assuming $V_{ji}=V_{ij}$, the
corresponding continuity equations are then given by
\begin{align}
\label{n4a}
   & \rho^\prime_i+3\mathcal{H}(1+w_i)\rho_i= Q_i,\\
    &Q_i= \sum_{j\neq i}V_{ij,\phi_j}(\phi_i,\phi_j)\phi_j^\prime\,.
        \label{n4b}    
\end{align}
Here the energy density $\rho_i$ and the pressure $p_i$ for the 
energy density of field $\phi_i$ are given by 
\begin{equation}
    \rho_i(p_i)=T_i\pm V_i(\phi_i)\pm\sum_{j\neq i}V_{ij}(\phi_i,\phi_j),\
        \label{n6}    
\end{equation}
where $T_i$ is the kinetic energy of field $\phi_i$, the $+$($-$) signs are for the cases $\rho_i$ ($p_i$),
and $w_i= p_i/\rho_i$.
The total energy density is then defined by
\begin{equation}
    \rho=\sum_i \rho_i-\sum_{i<j}V_{ij}(\phi_i,\phi_j),
        \label{n7}
        \end{equation}
where the last term on the right hand side is included to ensure no double counting. Eq.~(\ref{n7}) is a generalization of the two-field case~\cite{Aboubrahim:2024spa} to the case of $n$-interacting fields.  The observed relic density for the particle $i$ is given by 
\begin{align}
  \label{n8a}
   \Omega_{0i}&=\frac{\rho_i}{\rho_{0,\rm crit}}(1-\delta_i), 
   ~~~~~~\sum_i \Omega_{0i}=1,~~~~~~
\delta_i=\frac{\sum_{j\neq i} V_{ij}}{
 \sum_j \rho_j}.
\end{align}
Eqs.~(\ref{n4a}) and~(\ref{n4b}) are  a consistent set of continuity equations for the case of $n$ number of interacting fields, where the energy density $\rho_i$ corresponds to the field $\phi_i$.

\section{Field-theoretic inconsistency of the fluid model}

In order to include interactions, the phenomenological fluid model uses the continuity equations of the $\Lambda$CDM 
model but includes interactions by assuming a set of ad hoc
sources for them. 
To achieve conservation of energy, one then assumes that $\rho=\sum_i\rho_i$ and imposes the condition that the sum of the sources vanishes, i.e., $\sum_i Q_i=0$ which implies using Eq.~(\ref{n4a}) without knowledge of Eq.~(\ref{n4b}).
In field theory, $\sum_i Q_i=0$ leads to
\begin{align}
 V'_{\rm int}(\{\phi_i\},\{\phi_j\})&=0,
\label{vintp}
\end{align}
which means that $V_{\rm int}$ is a constant (independent of time). This is obviously incorrect since it depends on $n$ number of fields each of which is time-dependent and there is no reason why $V_{\rm int} (\phi_i,\phi_j)$ would be
independent of time.  We illustrate this with an example of two interacting fields $\chi$ and $\phi$ with an interacting potential 
\begin{align}
&V_{\rm int}(\phi,\chi)=\frac{\lambda}{2}\chi^2\phi^2,
\end{align}
one of which ($\chi$) will act as dark matter and the other ($\phi$) as dark energy or quintessence~\cite{Caldwell:2005tm,Pantazis:2016nky,Tsujikawa:2013fta}. In this case $Q_\chi$ and $Q_\phi$ are given by
\begin{align}
\label{rhocont}
&Q_\chi={V}_{\text{int},\phi}\phi_0^\prime,~~Q_\phi={V}_{\text{int},\chi}\chi_0^\prime.
\end{align}
From Eq.~(\ref{vintp}), we have
$V'_{\rm int}(\chi,\phi)=0$, i.e., $V_{\rm int}(\chi,\phi)$ is independent of time and so we write
\begin{align}
 V_{\rm int}(\chi,\phi)=\frac{\lambda}{2} \chi^2 \phi^2 = c. 
 \label{cte}
\end{align}
Here $c$ is a constant 
which gives $\chi=\pm \sqrt{\frac{2c}{\lambda}} \phi$, i.e., the fields $\chi$ and $\phi$  
are not independent but one is determined in terms of the other which is obviously 
false. Thus the fluid equations, Eq.~(\ref{eq2}) and Eq.~(\ref{eq3}), 
are inconsistent within a field theoretic framework.  A further explanation and deduction of consistent fluid equations for the case of two interacting fields is given in Appendix~\ref{appA}.

\section{Dark matter and dark energy evolution equations} 

In this section, we derive the background and perturbation equations that describe the time evolution of the DM and DE fields in a flat FLRW metric. In our interacting quintessence-dark matter model (QCDM), $\chi$ represents a dark matter 
field while $\phi$ is a dark energy field and are governed by the potentials
\begin{align}
\label{v1}
&V_1(\chi)=\frac{1}{2}m_\chi^2\chi^2, \\
\label{v2}
&V_2(\phi)=\mu^4\left[1+\cos\left(\frac{\phi}{F}\right)\right].
\end{align}
The interaction potential between $\chi$ and $\phi$ is based on field theory and is given by
\begin{align}
&V_{\rm int}(\phi,\chi)\equiv V_{12}(\phi,\chi)=\frac{\lambda}{2}\chi^2\phi^2.
\label{v3}
\end{align}
The fields $\chi$ and $\phi$ are perturbed, so that $\chi\to\chi+\chi_1$ and $\phi\to\phi+\phi_1$, where $\chi_1$ and $\phi_1$ are linear perturbations. The line element in the synchronous gauge is $\text{d}s^2=a^2(\tau)[-\text{d}\tau^2 +(\delta_{ij}+h_{ij})\text{d}x^i \text{d}x^j]$, where $h_{ij}$ is a metric perturbation.  

We track the evolution of the two background fields $\chi$ and $\phi$ by solving the Klein-Gordon (KG) equations 
\begin{align}
\label{kgc0}
&\chi^{\prime\prime}+2\mathcal{H}\chi^\prime+a^2(V_{1,\chi}+V_{12,\chi})=0, \\
&\phi^{\prime\prime}+2\mathcal{H}\phi^\prime+a^2(V_{2,\phi}+V_{12,\phi})=0,
\label{kgp0}
\end{align}
For a DM potential of the form given by Eq.~(\ref{v1}), the field undergoes rapid oscillations when $\mathcal{H}/m_\chi\ll 1$~\cite{Turner:1983he,Urena-Lopez:2015gur}, making a numerical solution intractable.  In order to efficiently solve the background evolution equation for
the DM field, we  recast Eq.~(\ref{kgc0}) in a different form. First, let us define the DM energy density as $\tilde{\rho}_\chi=\rho_\chi-V_{12}$ so that the modified energy density fraction becomes
\begin{align}
    \tilde{\Omega}_\chi\equiv\frac{\tilde{\rho}_\chi}{\rho_{\rm cr}}&=\frac{\kappa^2}{6\mathcal{H}^2}\Big(\chi^{\prime 2}_0+2a^2V_1(\chi_0)\Big),
    \label{om-tilde}
\end{align}
where $\kappa\equiv\sqrt{8\pi G}$. Based on Eq.~(\ref{om-tilde}), we define the following new dimensionless variables\cite{Copeland:1997et,Garcia-Arroyo:2024tqq}
\begin{align}
    \label{var-tilde-1}
    &\tilde{\Omega}_\chi^{1/2}\sin\left(\frac{\theta}{2}\right)=\frac{\kappa \chi^\prime}{\sqrt{6}\mathcal{H}}, \\
    \label{var-tilde-2}
    &\tilde{\Omega}_\chi^{1/2}\cos\left(\frac{\theta}{2}\right)=\frac{\kappa a V^{1/2}_{1}}{\sqrt{3}\mathcal{H}}, \\
    &y=-\frac{2\sqrt{2}\,a}{\mathcal{H}}\partial_\chi V^{1/2}_{1}.
    \label{var-tilde-3}
\end{align}
The advantage gained by introducing the new variables is that the rapid oscillations of the DM field can be absorbed into the $\theta$ variable. Using the new variables defined by Eqs.~(\ref{var-tilde-1})$-$(\ref{var-tilde-3}), we recast the KG equation of the field $\chi$ into a set of three coupled first order differential equations in $\Omega_\chi$, $\theta$ and $y$. After some lengthy, but otherwise straightforward, calculation, we arrive at
\begin{align}
    \Omega^\prime_\chi&=3\mathcal{H}\Omega_\chi(w_T-w_\chi) 
    +\frac{\kappa^2 a^2}{3\mathcal{H}^2}\Bigg[\mathcal{H}(1+3w_\chi)V_{12}-\chi^\prime V_{12,\chi}\Bigg],
    \label{omc0} \\
    \theta^\prime&=-3\mathcal{H}\sin\theta+\mathcal{H}y 
    -\frac{\kappa^2 a^2}{3\mathcal{H}^2\tilde{\Omega}_\chi}\Bigg(2\mathcal{H} V_{12}+\chi^\prime V_{12,\chi}\Bigg)\cot\frac{\theta}{2}, 
    \label{thetac0} \\
    y^\prime&=\frac{3}{2}{\cal H}(1+w_T)y\,,
    \label{yeq0}
\end{align}
with the total EoS $w_T=\sum p_i/\sum\rho_i$, where the sum is over all species (baryons, photons, neutrinos, DM and DE) and $\tilde{\Omega}_\chi=\Omega_\chi-(\kappa^2 a^2/3\mathcal{H}^2)V_{12}$. Note that the last set of terms in each of Eqs.~(\ref{omc0}) and~(\ref{thetac0}) represent the interaction term. In terms of the new variables, the DM equation of state $w_\chi$ can be written as
\begin{equation}
    w_\chi=\frac{\tilde{p}_\chi}{\tilde{\rho}_\chi}=\frac{\frac{1}{2a^2}\chi^{\prime 2}-V_1(\chi)}{\frac{1}{2a^2}\chi^{\prime 2}+V_1(\chi)}=-\cos\theta\,.
\end{equation}
With the rapid oscillations absorbed into the $\theta$ parameter, the DM equation of state oscillates between $+1$ and $-1$, so that the time-average of $w_\chi$ becomes zero, thus describing a pressureless fluid, i.e., CDM. We note here that the definition of the DM EoS is in terms of $\tilde p_\chi$ and $\tilde\rho_\chi$ which do not include the interaction term $V_{12}$. This is because the original dimensionless parameters, Eqs.~(\ref{var-tilde-1})$-$(\ref{var-tilde-3}), are defined in terms of $V_1$ only. This allows for a more straightforward derivation of the DM background equations. The interaction term, however, appears in the evolution equations, Eqs.~(\ref{omc0})$-$(\ref{yeq0}). 

For linear perturbations, we solve the KG equations of $\chi_1$ and $\phi_1$ which are given by
\begin{align}
    \label{kgp1}
    \phi_1^{\prime\prime}+2\mathcal{H}\phi_1^\prime+(k^2+a^2V_{,\phi\phi})\phi_1+a^2V_{,\phi\chi}\chi_1+\frac{1}{2}h^\prime\phi^\prime=0, \\
    \chi_1^{\prime\prime}+2\mathcal{H}\chi_1^\prime+(k^2+a^2V_{,\chi\chi})\chi_1+a^2V_{,\chi\phi}\phi_1+\frac{1}{2}h^\prime\chi^\prime=0,
    \label{kgc1}
\end{align}
where $h$ is the trace of $h_{ij}$. We then determine the energy density and pressure perturbations of the fields using
\begin{align}
    \delta\rho_\phi&=\frac{1}{a^2}\phi^\prime\phi_1^\prime+({V}_2+{V}_{12})_{,\phi}\phi_1+{V}_{12,\chi}\chi_1, \\
    \delta p_\phi&=\frac{1}{a^2}\phi^\prime\phi_1^\prime-({V}_2+{V}_{12})_{,\phi}\phi_1-{V}_{12,\chi}\chi_1, \\
    \delta\rho_\chi&=\frac{1}{a^2}\chi^\prime\chi_1^\prime+({V}_1+{V}_{12})_{,\chi}\chi_1+{V}_{12,\phi}\phi_1, \\
    \delta p_\chi&=\frac{1}{a^2}\chi^\prime\chi_1^\prime-({V}_1+{V}_{12})_{,\chi}\chi_1-{V}_{12,\phi}\phi_1.
\end{align} 
The velocity divergence $\Theta=ik^i v_i$ of the fields are calculated as
\begin{align}
    (\rho_\phi+p_\phi)\Theta_\phi&=\frac{k^2}{a^2}\phi^\prime\phi_1, \\
    (\rho_\chi+p_\chi)\Theta_\chi&=\frac{k^2}{a^2}\chi^\prime\chi_1\,.
\end{align}
The density contrast is then defined as $\delta=\delta\rho/\rho$.

\section{Constraints from background cosmology on interacting dark matter and dark energy} 

We are now in a position to solve Eqs.~(\ref{omc0}),~(\ref{thetac0}) and~(\ref{yeq0}) to track the evolution of the DM field $\chi$, and the KG equation, Eq.~(\ref{kgp0}), to determine the evolution of the DE field.
We implement and solve the above equations in the Boltzmann equation solver \code{CLASS}~\cite{Blas:2011rf} in order to determine the evolution of the DM and DE fields as well as the SM species (photons, baryons and neutrinos).  
To make sure the closure relation $\sum_i \Omega_i=1$ is satisfied, we use the shooting method in \code{CLASS} during the evolution of the background equations starting from $a_{\rm ini}=10^{-14}$ to $a_0=1$ (today). The shooting procedure is used by \code{CLASS} to determine the initial values of parameters at $a_{\rm ini}$ so that the background evolution produces results that match desirable values today (at $a_0=1$). For example, we use in our analysis the attractor initial conditions~\cite{Urena-Lopez:2015gur} for the field $\chi$ with
\begin{align}
    y_{\rm ini}&=\frac{2m_\chi}{H_0}a_{\rm ini}^2\Omega^{-1/2}_{\text{rad}_0}, \\
    \theta_{\rm ini}&=\frac{1}{5}y_{\rm ini}, \\
    \Omega_{\chi_{\rm ini}}&=a_{\rm ini}\frac{\Omega_{\chi_0}}{\Omega_{\text{rad}_0}}\left[\frac{4\theta_{\rm ini}^2}{\pi^2}\left(\frac{9+\pi^2/4}{9+\theta_{\rm ini}^2}\right)\right]^{3/4},
\end{align}
where the index `0' indicates a quantity's value today and `rad' stands for radiation. However, to determine the value of $\Omega_{\chi_{\rm ini}}$ that produces that desired value of $\Omega_{\chi 0}\sim 0.26$ today, we include a small shooting parameter $s\sim 10^{-2}$ so that
\begin{equation}
   \Omega_{\chi_{\rm ini}} \to s+\Omega_{\chi_{\rm ini}}, 
\end{equation}
and allow \code{CLASS} to vary $s$ in order to achieve the imposed tolerance. For the DE field $\phi$, the fractional energy density today is determined by \code{CLASS} to match the budget equation so that $\Omega_{\phi 0}^{\rm target}=1-\sum_i\Omega_i$, where the sum runs over all species (except DE). Now the shooting algorithm needs to determine $\phi_{\rm ini}$ so that $|\Omega_{\phi 0}(\phi_{\rm ini})-\Omega_{\phi 0}^{\rm target}|<\epsilon$, where $\epsilon\sim 10^{-2}$ and $\phi_{\rm ini}$ stands as the shooting parameter. The background equations are integrated from $a_{\rm ini}$ to $a_0$ and the obtained fractional density of DE is checked against the target value. If the required tolerance is not reached, \code{CLASS} adjusts $\phi_{\rm ini}$ and starts over until convergence is achieved. The DE field velocity is chosen as $\phi^\prime_{\rm ini}=10^{-3}$ and an estimate of $\mu^4$ is determined by minimizing the DE potential so that $\mu^4=\frac{3}{2}H_0^2\Omega_{\phi 0}$. This value serves as an initial estimate and we modify $\mu^4$ accordingly in order to achieve a consistent cosmology.

\subsection{The strong coupling regime}

The DM-DE coupling strength, $\lambda$, enters Eqs.~(\ref{kgp0}),~(\ref{omc0}) and~(\ref{thetac0}) through the term $V_{12}$ and affects the DE EoS through
\begin{equation}
    w_\phi\equiv\frac{p_\phi}{\rho_\phi}=\frac{\frac{1}{2a^2}\phi^{\prime 2}+V_2+V_{12}}{\frac{1}{2a^2}\phi^{\prime 2}-V_2-V_{12}}\,.
\end{equation}
Before going in depth into analyzing the effect of $\lambda$ on $w_\phi$, we discuss some of the important observables and their evolution for three values of $\phi_{\rm ini}$ and $\mu^4$ (expressed in standard \code{CLASS} units of $m_{\rm pl}$ and $m_{\rm pl}^2/\mathrm{Mpc}^2$, respectively) with non-zero DM-DE interaction, i.e., $\lambda>0$. Here we choose $\lambda>10^{-2}$ which corresponds to the strong coupling regime. 

\begin{figure}[H]
\begin{centering}
\includegraphics[width=0.49\linewidth]{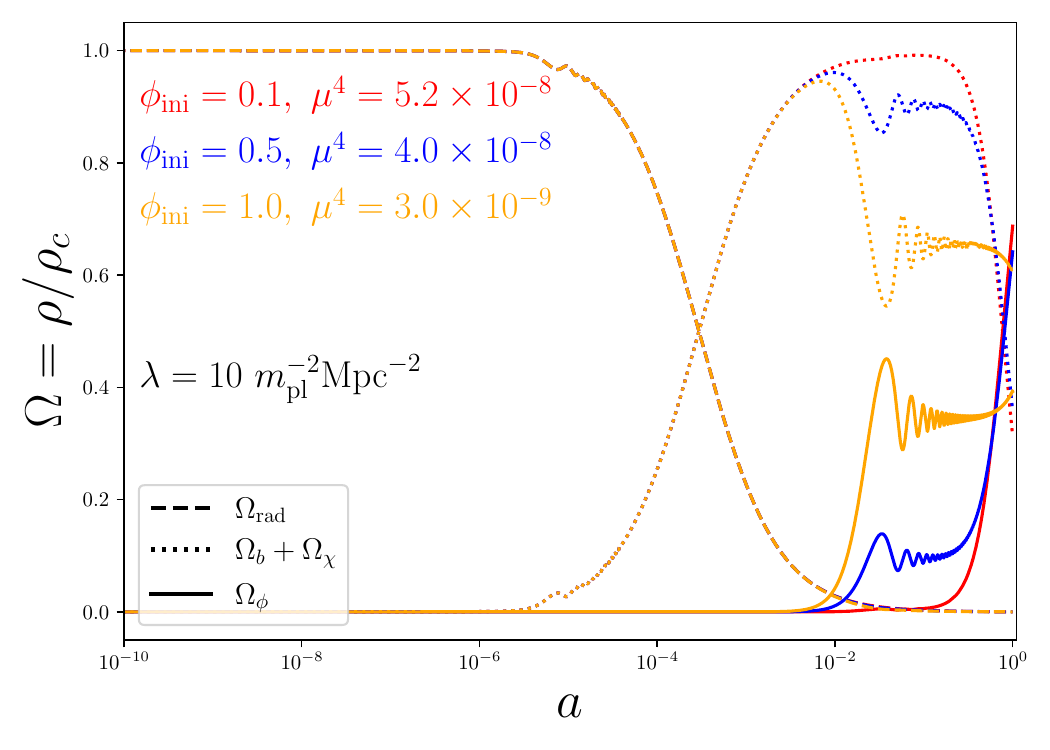}
\includegraphics[width=0.49\textwidth]{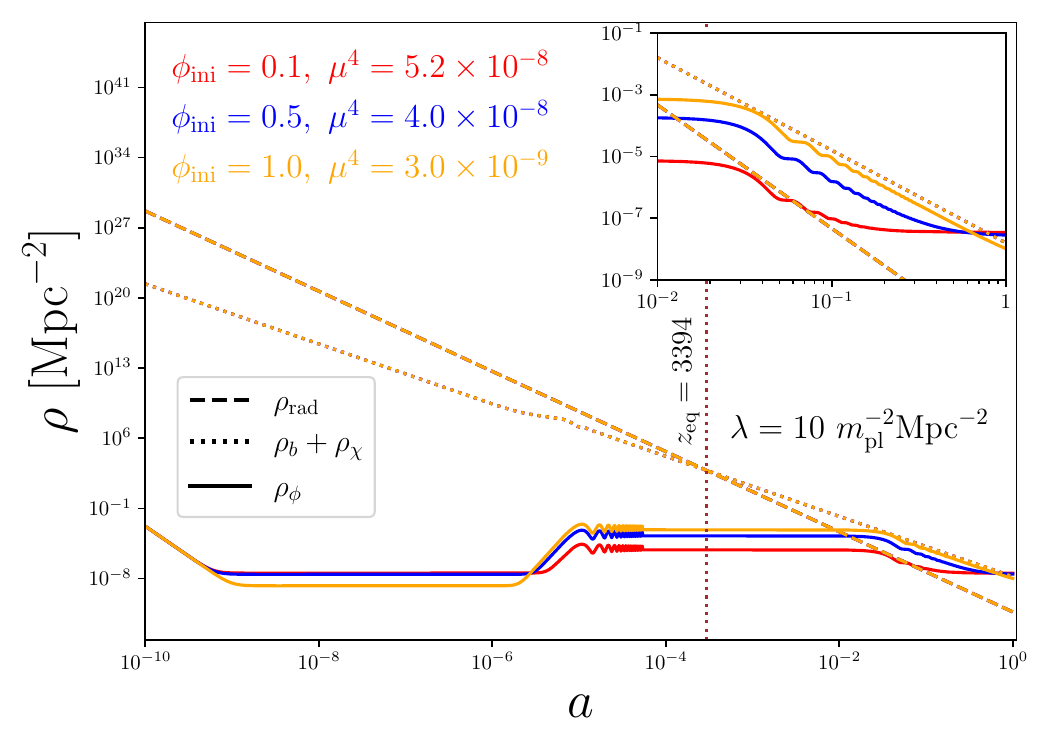} \\
\includegraphics[width=0.49\linewidth]{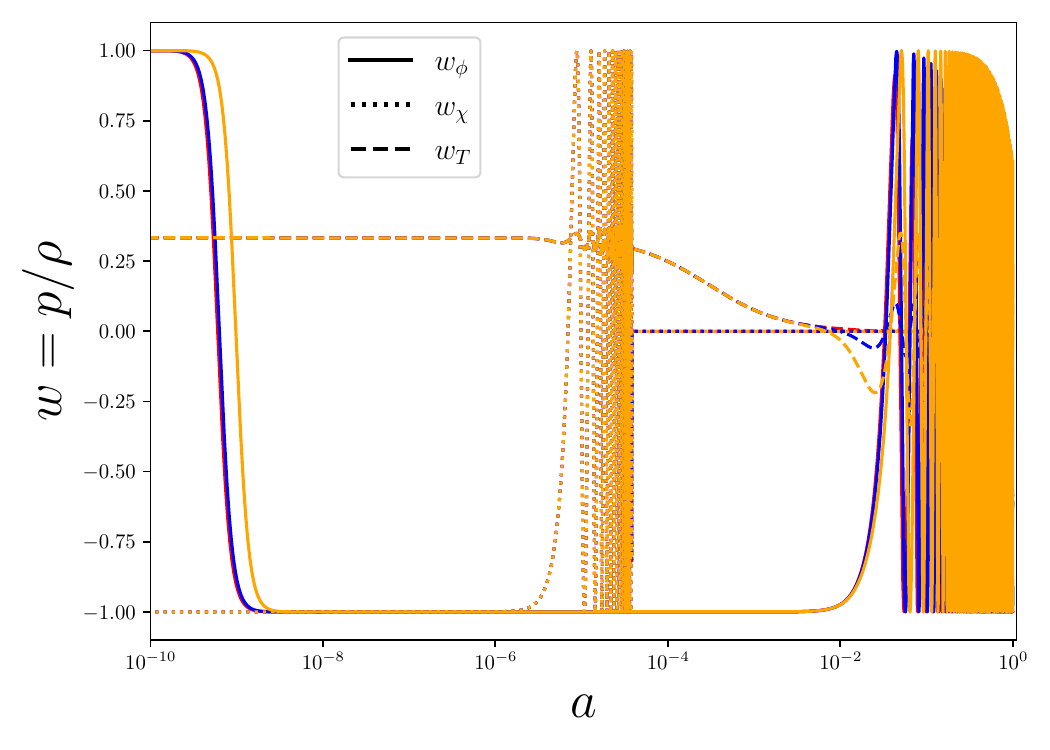}
\includegraphics[width=0.49\textwidth]{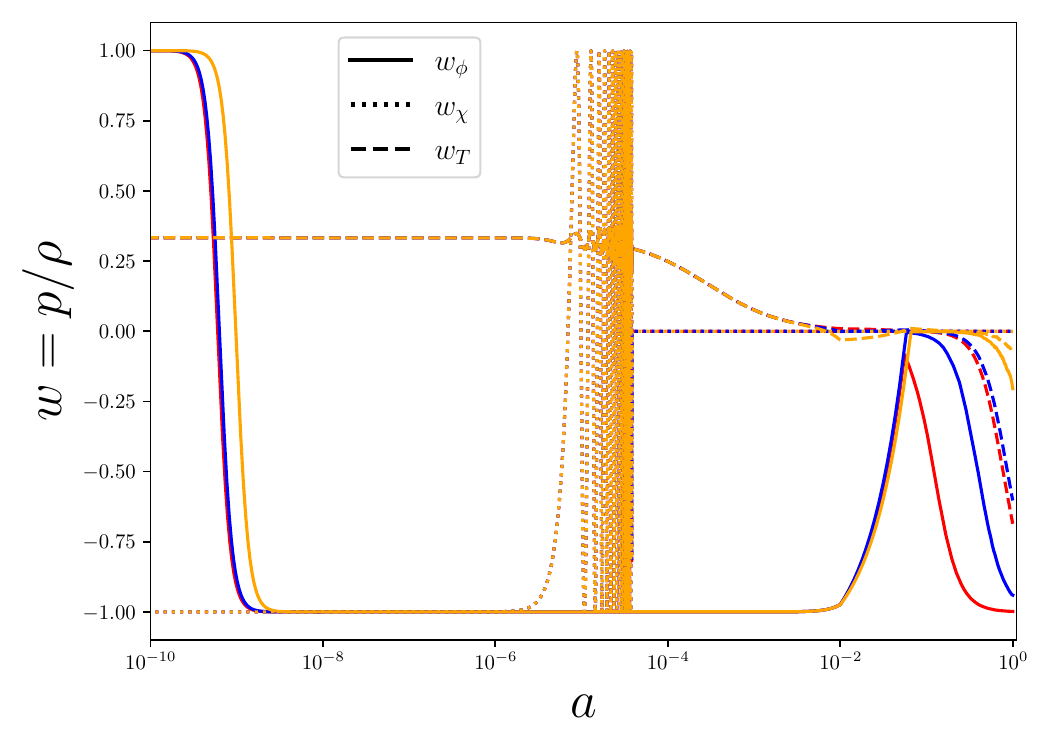}
\caption{Top panels: plot of the energy density fractions (left) and the energy densities evolution (right) as a function of the scale factor $a$. Bottom panels: plot of the evolution of the equations of state (EoS) of DM and DE and the total EoS (left) and a time-average over the oscillations in $w_\phi$ and $w_T$ (right panel). The DM EoS, $w_\chi$, is being averaged over in both panels after a short period of oscillations. The color code in the top panel is the same for the bottom panel and correspond to different values of $\phi_{\rm ini}$ and $\mu^4$ in the presence of DM-DE interaction, expressed in standard \code{CLASS} units of $m_{\rm pl}$ and $m_{\rm pl}^2/\mathrm{Mpc}^2$, respectively.}
\label{fig1}
\end{centering}
\end{figure}

\begin{figure}[H]
\begin{centering}
\includegraphics[width=1.0\linewidth]{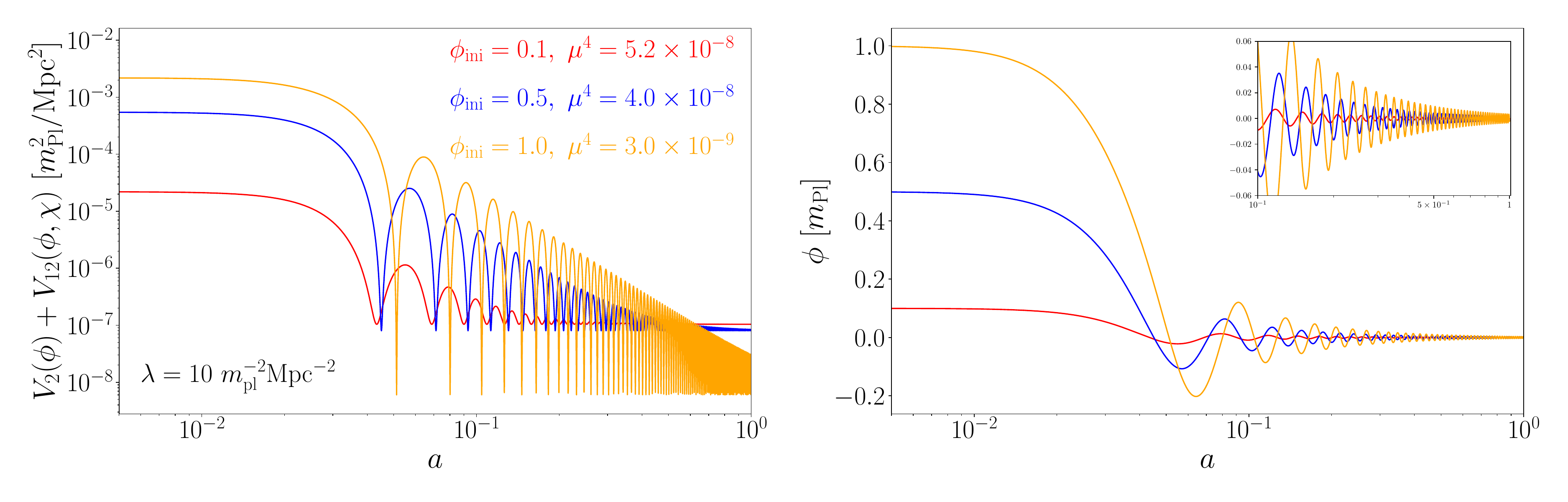}
\caption{Evolution of the total potential $V_2(\phi)+V_{12}(\phi,\chi)$ (left panel) and the scalar DE field $\phi$ (right panel) against the scale factor $a$ in the presence of DM-DE interaction. The three benchmarks for $(\phi_{\rm ini},\mu^4)$ are the same as in Fig.~\ref{fig1}. The inset in the right panel gives a close up view of the oscillations in the DE field. }
\label{fig1a}
\end{centering}
\end{figure}

In the top panel of Fig.~\ref{fig1}, we exhibit the evolution of the energy density fraction (left) and of the energy density (right) for baryons, photons, DM and DE. Similar to $\Lambda$CDM, the QCDM model predicts the redshift of matter-radiation equality to be $z_{\rm eq}\sim 3390$ consistent with Planck's measurements. Larger values of $\mu^4$ lead to larger $\Omega_{\phi 0}$ and thus smaller $\Omega_{\chi 0}$ so that the closure relation holds. To see this, we examine the evolution of the DE scalar field $\phi$ and the potential terms in Fig.~\ref{fig1a}, where the left panels shows the total potential $V_2+V_{12}$ vs. $a$ and the right panel tracks the field $\phi$ vs. $a$. The plots clearly show an oscillatory behavior of the scalar field at late times and the total potential attaining a larger value at $a=1$ for larger $\mu^4$ values. This explains why $\Omega_{\phi 0}$ is larger for higher $\mu^4$. Note that different $\phi_{\rm ini}$ values have no effect on $\phi$ today. 
The oscillatory feature in $\Omega$ can be explained by examining the bottom left panel which shows the equation of state (EoS) of DM, DE and the total EoS.  At early times, $w_\chi=-1$ and $w_\phi=+1$ and so the $\phi$ field acts as radiation while the $\chi$ field acts as an early DE component for a period of time before the onset of rapid oscillations at around $a\sim 10^{-5}$. Averaging the fast oscillations in $\chi$ renders $w_\chi=0$ and so the $\chi$ field eventually dilutes as CDM, i.e., $\rho_\chi\sim a^{-3}$. The imprints of these rapid oscillations are visible as little wiggles in the energy density plots in the left-hand-side of the top panel. The $\phi$ field, on the other hand, starts off as radiation and as the DE potential rolls down to its minimum, the EoS drops to $-1$. The interesting feature here is that similar to $\chi$, the DE field begins to oscillate but at a later time, around $a\sim 10^{-2}$. The difference between the oscillations of the $\chi$ and $\phi$ fields is that while the amplitude of oscillations of $w_\chi$ is constant, that of $\phi$ is decaying. For $\phi/F\ll 1$, the KG equation of the $\phi$ field, Eq.~(\ref{kgp0}), takes the form
\begin{equation}
\phi^{\prime\prime}+2\mathcal{H}\phi^\prime+a^2m_\phi^2\,\phi\approx 0, 
\end{equation}
where $m^2_\phi=\lambda\chi^2-\mu^4/F^2$. For $\lambda$ large enough such that $m_\phi^2>0$, oscillations in the DE field begin once $\mathcal{H}/m_\phi\ll 1$. If one examines $\rho_\phi$ closely in the upper right panel of Fig.~\ref{fig1}, it is clear that there are two oscillatory features over two time scales. The first is in the range $10^{-5}<a<10^{-4}$ which comes from oscillations in the DM field and that is carried over to DE via the coupling $\lambda$. The second is around $a\sim 0.1$, visible in the inset of Fig.~\ref{fig1}, and this comes directly from solutions to the KG equation when $\mathcal{H}/m_\phi\ll 1$. 
Averaging over these oscillations, we display in Fig.~\ref{fig1} (bottom right panel) the variations in $w_\phi$ where one can see $w_\phi$ departing away from $-1$ before turning around and decreasing again toward $-1$. One finds that for $\phi_{\rm ini}=0.1$, $w_\phi$ decays to $-1$ as one would expect from a DE component, while for $\phi_{\rm ini}=0.5$ we get $w_\phi\sim -0.9$. However, for $\phi_{\rm ini}=1.0$, the oscillations remain strong even at $a=1$ and even though the averaged $w_\phi$ is less than zero, it is still much larger than $-1$, which is the well accepted value from observations. Therefore, one can easily reject $\phi_{\rm ini}=1.0$ and only consider smaller $\phi_{\rm ini}$. Note that even when the oscillations in $\Omega_\chi$ have been averaged out, they reappear once $\Omega_\phi$ starts its oscillations because the fields are coupled. One final remark concerning oscillations pertains to an apparent small oscillatory feature which is also visible in $\Omega_{\rm rad}$. This is not arising from a possible oscillation in $\rho_{\rm rad}$ (the right panel of Fig.~\ref{fig1} clearly shows that), but rather from $\rho_{c}$ (critical density) which is obtained from the sum of all energy densities of the species. Therefore, $\rho_c$ inherits any oscillation in $\rho_\chi$ and $\rho_\phi$ and transfers it to $\Omega_{\rm rad}$.

\begin{figure}[H]
\begin{centering}
\includegraphics[width=0.6\textwidth]{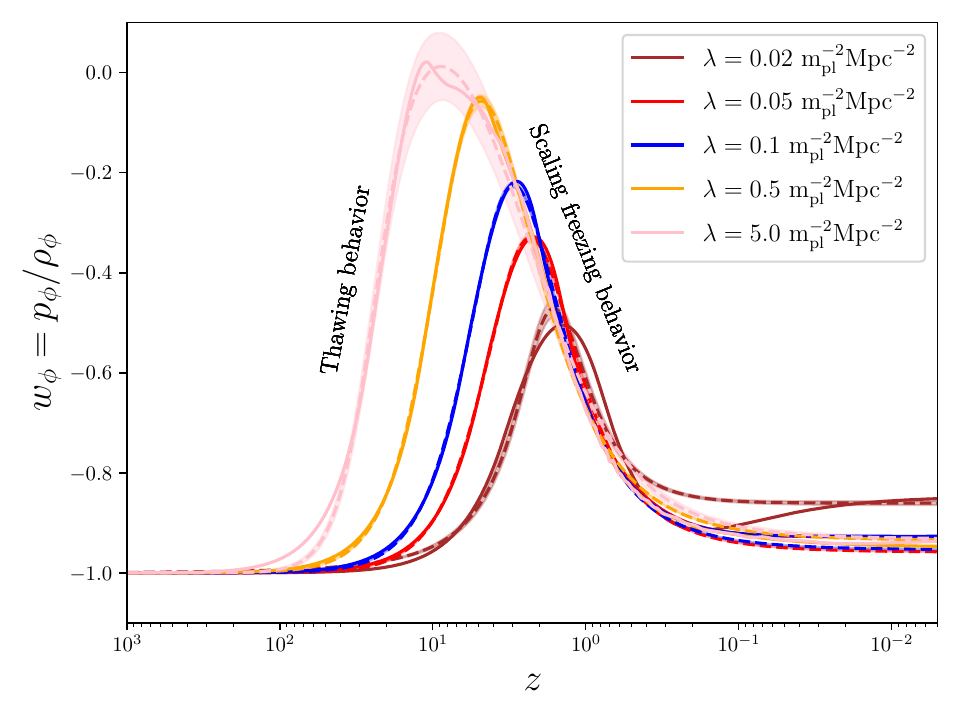} 
\caption{Evolution of $w_\phi$ for quintessence for five values of $\lambda$ as a function of the redshift $z$. The solid lines are $w_\phi$ predicted by QCDM and the dashed lines are fits based on  Eq.~(\ref{fit}) including a  $1\sigma$ error band around each curve. In each case, quintessence transmutes from thawing (left shoulder) to scaling freezing (right shoulder) induced by $\lambda$.}
\label{fig2}
\end{centering}
\end{figure}

We now address the peculiar evolution of $w_\phi$ seen in the lower right panel of Fig.~\ref{fig1} for $a>10^{-2}$. We show in Fig.~\ref{fig2} the evolution of the DE equation of state as a function of the redshift $z=1/a-1$ for five values of the interaction strength $\lambda$ (solid curves) with the dashed curves representing the fits to the data from the QCDM model. The fit is given by
\begin{equation}
    w(a)=-1+\frac{\alpha \,a^p e^{-p\,a}}{1+(\beta a)^q},
    \label{fit}
\end{equation}
where $\alpha$, $\beta$, $p$ and $q$ are the fit parameters. It is clear that for the considered benchmarks, the EoS departs away from $w_\phi=-1$ in the range $10<z<100$ tracking a thawing quintessence behavior, before turning around and falling back toward $w_\phi\sim -1$ at late times, a behavior similar to scaling freezing. The presence of a DM-DE interaction term has induced a transmutation of the DE from thawing to scaling freezing. A quintessence with a scaling freezing behavior at current times is not favored by the recent DESI data. 

\begin{figure}[H]
\begin{centering}
\includegraphics[width=0.49\textwidth]{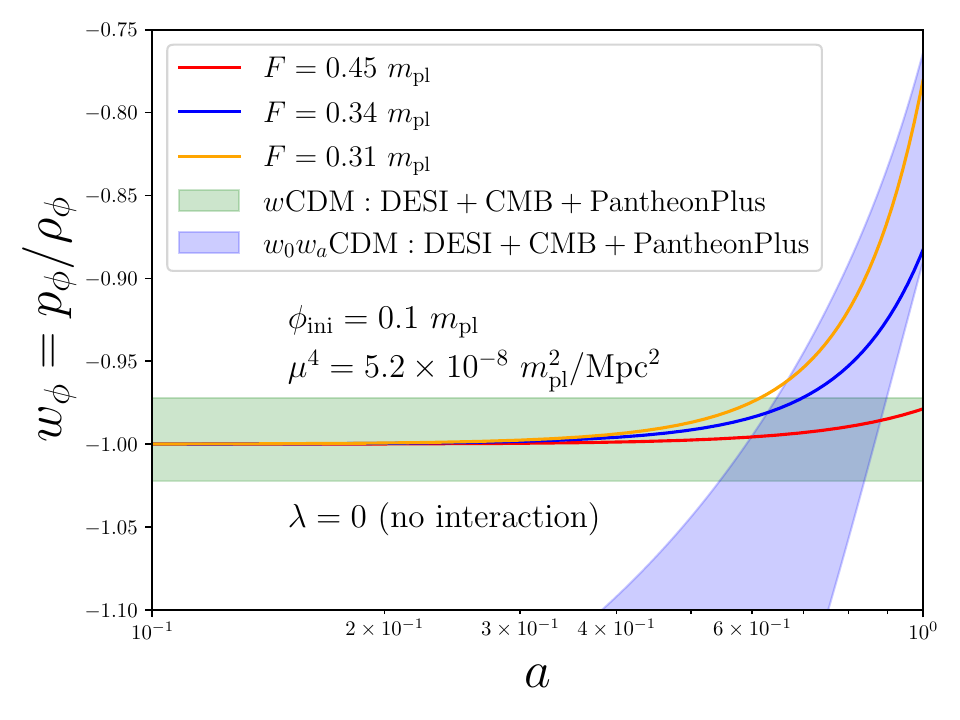}
\includegraphics[width=0.49\textwidth]{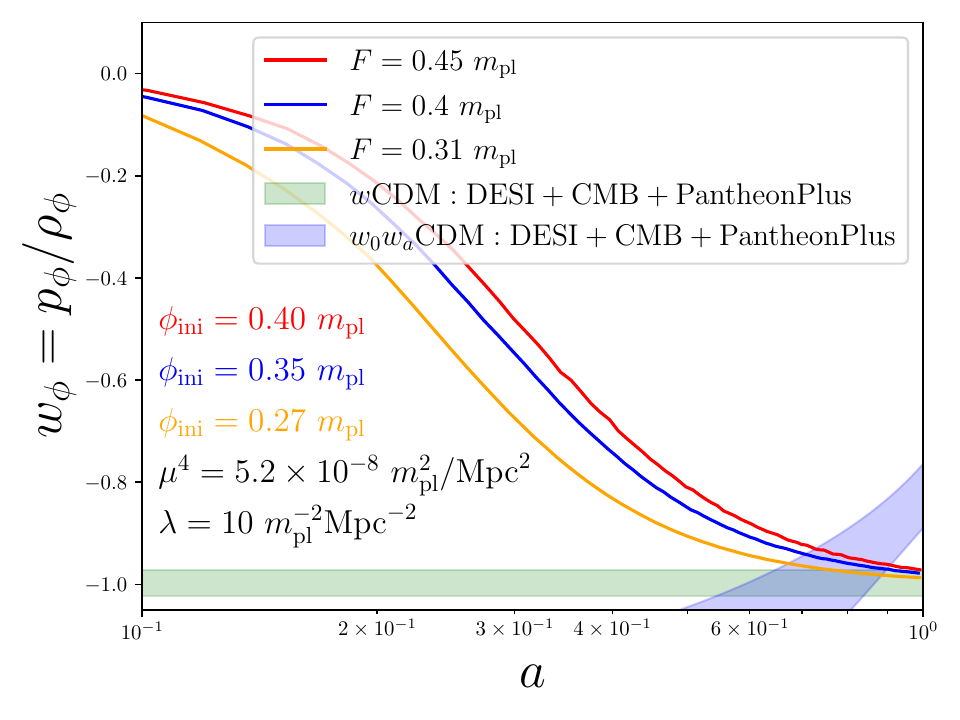}
\caption{A plot of the EoS of $\phi$ as a function of the scale factor $a$ exhibiting the thawing behavior (left) when $\lambda=0$ and scaling freezing (right) when $\lambda\neq 0$. The $1\sigma$ regions from DESI are shown as blue and green bands. A comparison of the left and the right panels indicates that the EoS of dark energy is sensitively dependent on the DE-DM interaction strength. }
\label{fig3}
\end{centering}
\end{figure}

To see this, we examine two cases: when no DM-DE interaction is presented and when the interaction is switched on. The first scenario is shown in the left panel of Fig.~\ref{fig3} and the second in the right panel of Fig.~\ref{fig3}, where we exhibit the evolution of the DE EoS between $a=0.1$ and $a=1$. The green band shows the $1\sigma$ range of the allowed values of $w_{\phi}$ as obtained from the combination of the data set $\mathrm{DESI}+\mathrm{CMB}+\mathrm{PantheonPlus}$ under the $w$CDM assumption 
($w_\phi=\text{constant}$), which is consistent with the baseline $\Lambda$CDM. The blue band corresponds to DESI's $1\sigma$ values of $w_0$ and $w_a$ in the CPL parametrization, given by $w(a)=w_0+(1-a)w_a$ (dubbed $w_0w_a$CDM).

In the absence of DM-DE interaction (left panel), the EoS tracks $w_\phi=-1$ before it starts to deviate away from it close to $a=0.3$ which is consistent with thawing models with a pseudo-Nambu-Goldstone boson potential~\cite{Frieman:1995pm}. The benchmark in red is consistent with the DESI $w$CDM result\footnote{Fits to DESI results have been carried out recently for quintessence, e.g. in refs.~\cite{Tada:2024znt,Reboucas:2024smm,Berghaus:2024kra,Wolf:2024stt,Gialamas:2024lyw}.} (green band), while the other two benchmarks show major deviation away from $w_\phi=-1$ and are consistent with $w_0w_a$CDM near $a=1$. When the DM-DE interaction is switched on (right panel), we notice that the dynamics of the EoS changes significantly. The evolution of $w_\phi$ becomes consistent with scaling freezing models~\cite{Ferreira:1997au,Copeland:1997et}, where the EoS slowly approaches $w_\phi=-1$ as $a\to 1$. Therefore, the presence of the interaction term has modified the DE potential in such a way that the new behavior now resembles that of scaling freezing models with an evolution that fits well a double exponential potential $V_S(\phi)=\tilde{V}_0(e^{-\lambda_1 \phi}+e^{-\lambda_2 \phi})$. Note that both the thawing and scaling freezing modes pertaining to $w_\phi$ appear in the lower right panel of Fig.~\ref{fig1} where there is a clear transition from thawing to scaling freezing near $a_t\sim 0.1$ for non-zero interaction. Analytically, one can see this behavior by expanding the potentials around the minimum, so that 
\begin{equation}
 V_2(\phi)+V_{12}(\phi,\chi)=\mu^4\left[1+\cos\left(\frac{\phi}{F}\right)\right]+\frac{\lambda}{2}\chi^2\phi^2\simeq 2\mu^4+V_0\phi^2,  
 \label{vs1}
\end{equation}
where $V_0=(\lambda\chi^2/2-\mu^4/2F^2)$. Fitting the QCDM model data to the double exponential potential, we find that $\lambda_1=-\lambda_2$, and so 
\begin{equation}
 V_S(\phi)=\tilde{V}_0(e^{-\lambda_1 \phi}+e^{\lambda_1 \phi})=2\tilde{V}_0\cosh(\lambda_1\phi)\simeq 2\tilde{V}_0+\tilde{V}_0\lambda_1^2\phi^2\,.   
 \label{vs2}
\end{equation}
Notice that Eqs.~(\ref{vs1}) and~(\ref{vs2}) describe the same physics which explains the transition between thawing and scaling freezing quintessence in the late universe.

\begin{figure}[H]
    \centering
    \includegraphics[width=0.49\linewidth]{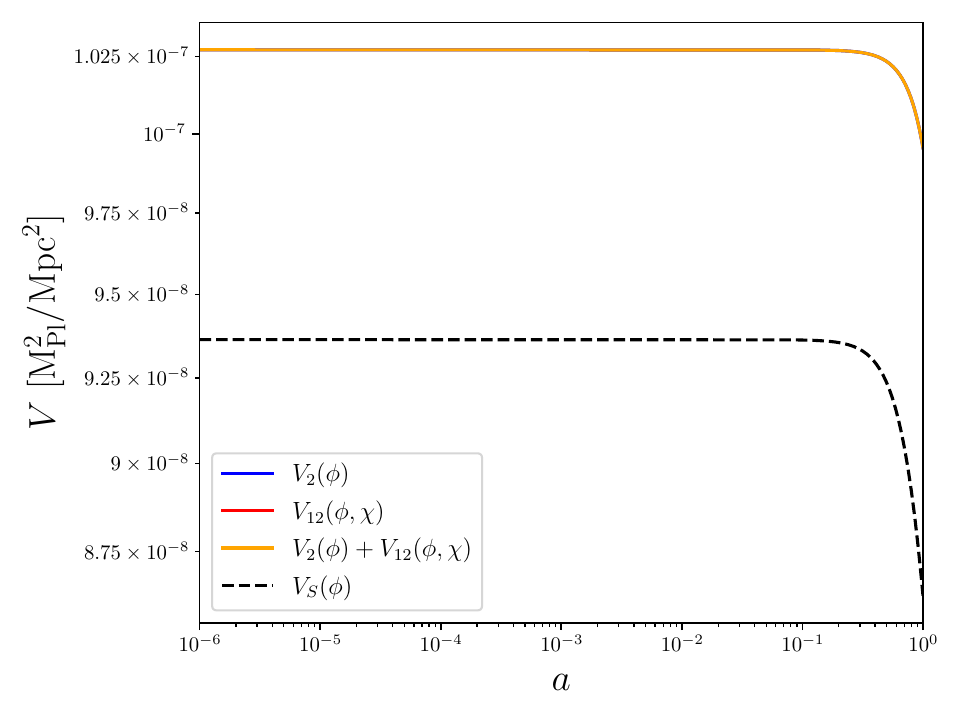}
    \includegraphics[width=0.49\linewidth]{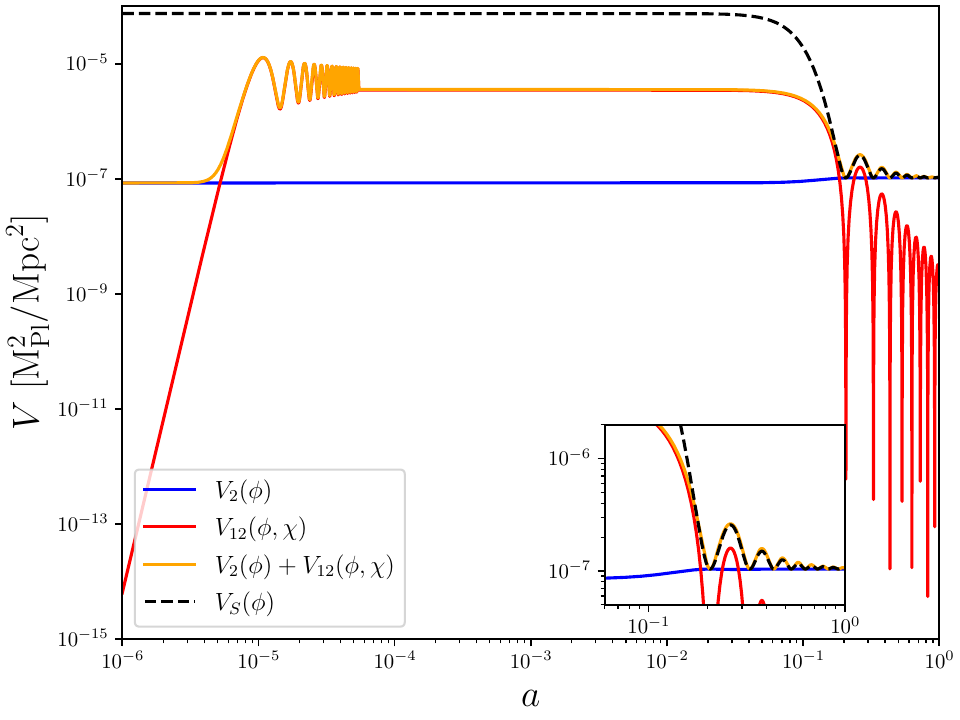}
    \caption{\normalsize Left panel versus right panel correspond to the no interaction versus non-zero interaction cases. The evolution of the quintessence potential $V_2(\phi)$ (blue), the interaction potential $V_{12}=\lambda \phi^2\chi^2/2$ (red) and the total potential (orange) as a function of the scale factor $a$. The potential $V_S(\phi)=\tilde{V}_0(e^{-\lambda_1 \phi}+e^{-\lambda_2 \phi})$ (black dashed) is fitted to the total potential, where $\lambda_1=-\lambda_2=-18.2\,\,\text{m}_{\rm pl}^{-1}$. The fit works very well for $a>0.1$ (inset of right panel). }
    \label{fig4}
\end{figure}

In Fig.~\ref{fig4}, we show a plot for the potential versus $a$ when no DM-DE interaction is present (left panel) and when the interaction is switch on (right panel). The total potential shown as an orange curve in the right panel has a decaying oscillatory feature close to $a=1$ and a double exponential function, $V_S(\phi)$, shown as a black dashed curve, exactly fits the total potential in that region, which can also be clearly seen in the figure inset. There, $V_2+V_{12}\sim \phi^2$ which is exactly the behavior of the scaling freezing double exponential potential for $\lambda_1=-\lambda_2$.  

Based on DESI DR1 and DR2, a DE quintessence with a freezing behavior is inconsistent with the existing constraints on $w_0$ and $w_a$. Our model, however, has several free parameters, namely: $F$, $\phi_{\rm ini}$, $\mu^4$ and $\lambda$. We thoroughly checked that no parameter combination can be chosen that will fit the $w_0$ and $w_a$ values in the strong coupling regime. We did so by expanding Eq.~(\ref{fit}) around $\epsilon=(1-a)$ up to linear order. We get
\begin{align}
   \label{w0}
   w_0&=-1+ \frac{\alpha e^{-p}}{1+\beta^q},\\
   w_a&= \frac{q \beta^q}{1+ \beta^q} (1+w_0)\,.
   \label{wa}
\end{align}
From the above we see that the current DESI analysis fixes two of the four parameters  of Eq.~(\ref{fit}).
Thus, for example, one may use $w_0$ and $w_a$ to determine $\beta^q$ using Eq.~(\ref{wa}) and use 
Eq.~(\ref{w0}) to determine $e^{-p}$. This leaves us with two unconstrained parameters $\alpha$ and $q$. Knowing the values of $w_0$ and $w_a$ from DESI, we tried to fit our model data using $\alpha$ and $q$ and found that one cannot get any good fit in the strong coupling regime. As expected, the current DESI data does not favor quintessence with a scaling freezing behavior.

\subsection{The weak coupling regime}

We have seen in the previous section that a strong DM-DE coupling turns a thawing quintessence into freezing at late time which is problematic from an observational perspective. In the absence of a DM-DE interaction, quintessence with potential of the form given by Eq.~(\ref{v2}) give rise to a thawing behavior as depicted in the left panel of Fig.~\ref{fig3}. In this section we will answer the following questions: (1) How large can the DM-DE interaction strength, $\lambda$, be while still remaining consistent with DESI's data within $2\sigma$, and (2) does any of the data sets in DESI's analysis prefer a non-zero $\lambda$? We approach the latter question by investigating the impact of $\lambda$ and the other three input parameters: $F$, $\phi_{\rm ini}$ and $\mu^4$ in driving the $w_0$ and $w_a$ values to fall within DESI's $2\sigma$ bounds. 

\begin{figure}[H]
\begin{centering}
\includegraphics[width=0.49\textwidth]{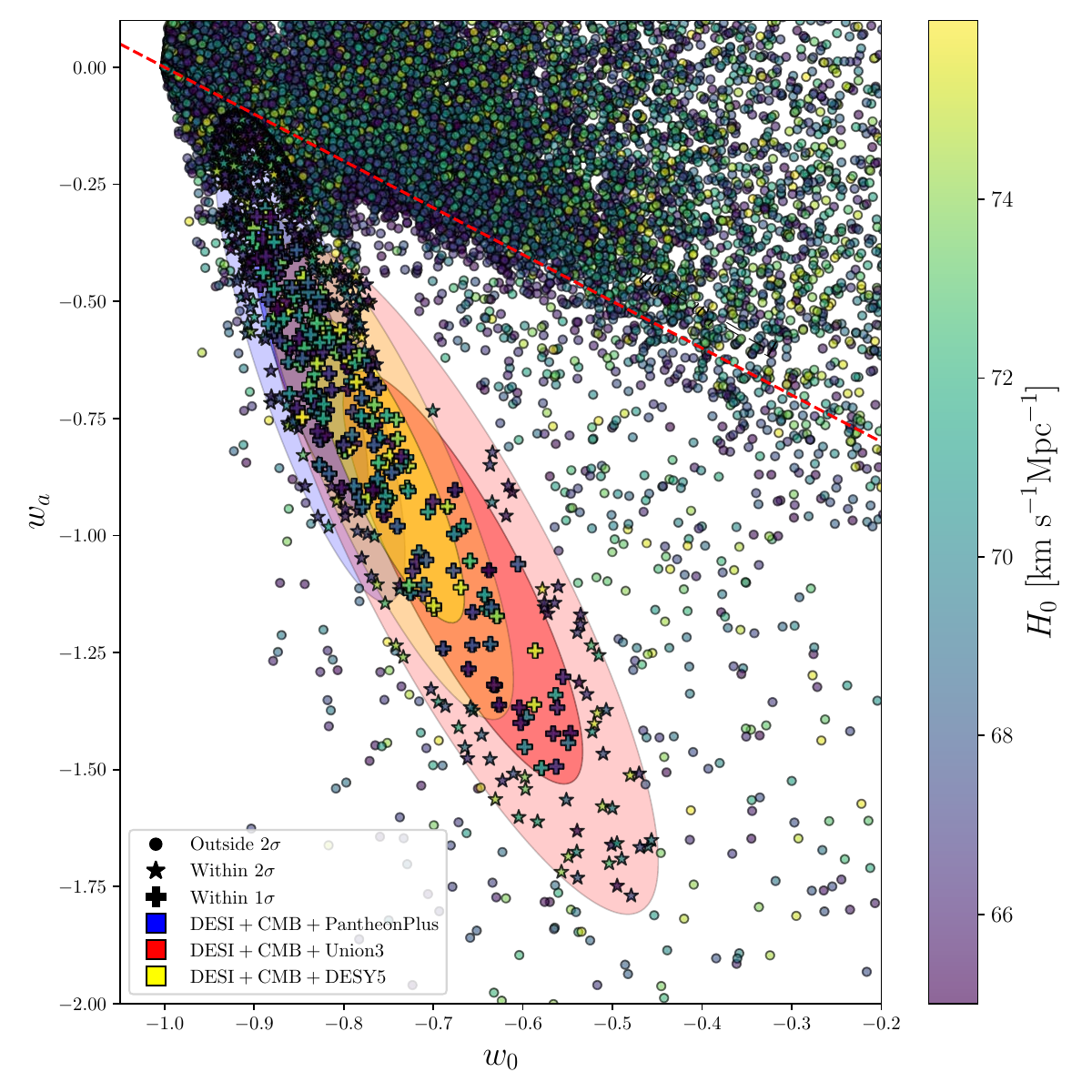}
\includegraphics[width=0.49\textwidth]{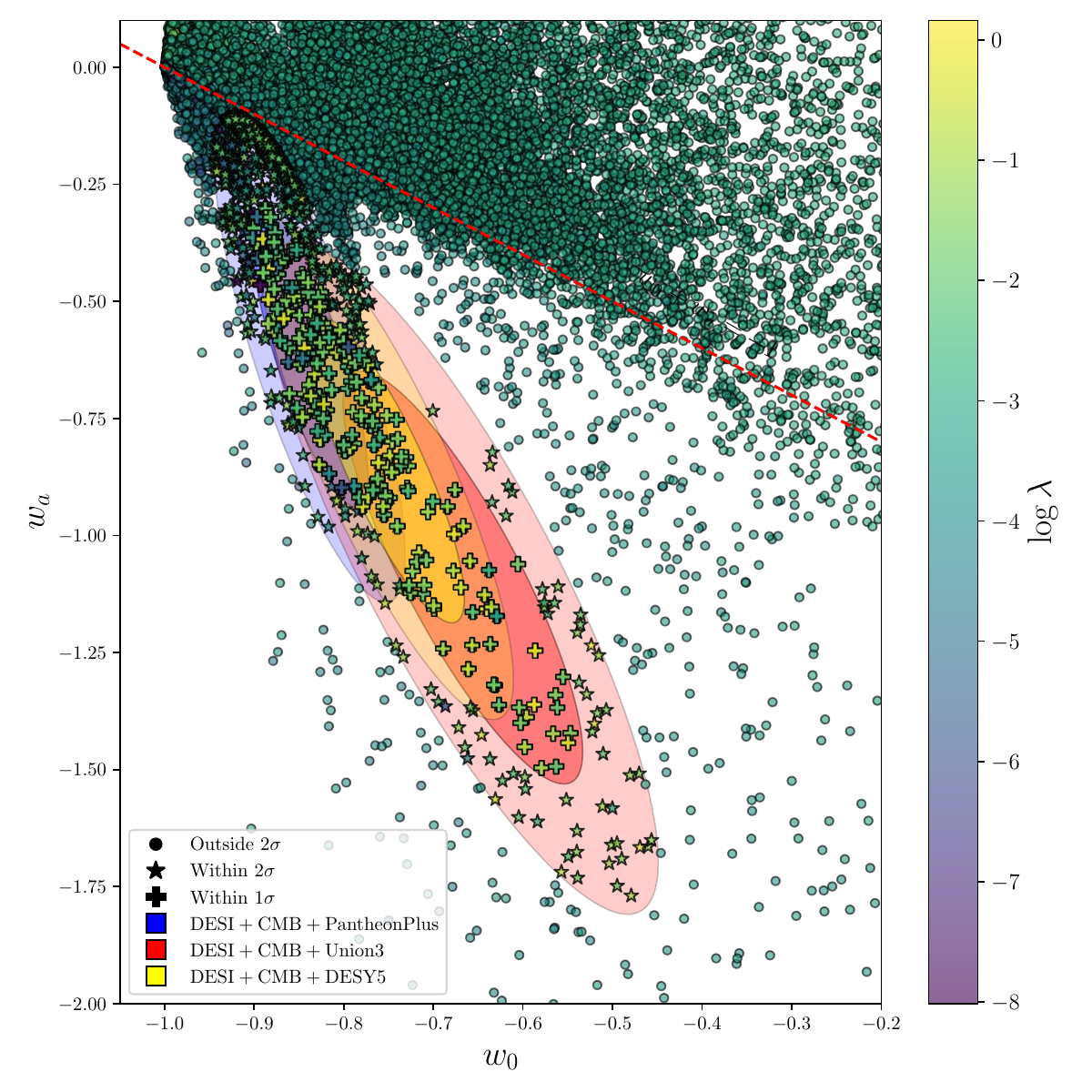}
\caption{Scatter plots in the $w_0$-$w_a$ plane overlaid on top of the posterior contours from DESI DR2. The left panel exhibits the Hubble parameter $H_0$ as the color axis while the right panel shows $\log\lambda$. The model is able to produce benchmarks that lie in the $1\sigma$ and $2\sigma$ posteriors of the considered data sets. }
\label{fig5}
\end{centering}
\end{figure}

We run a MCMC analysis of the model parameter space while restricting the sampling to the weak coupling regime, i.e., $\lambda\leq 10^{-2}$. We find that in this regime, the dark energy EoS can be fitted by the function
\begin{align}
    \hat{w}(a)&=-1+\alpha\,e^{-\beta\,a}\arctan(p\,a^q),
    \label{fit-weak}
\end{align}
with $\alpha$, $\beta$, $p$ and $q$ being the fit parameters. Expanding Eq.~(\ref{fit-weak}) up to first order in $(1-a)$, we get $w_0$ and $w_a$ so that
\begin{align}
\label{w0-weak}
    w_0&=-1+\alpha\,e^{\beta}\,\arctan(p)\,, \\
    w_a&=\frac{\alpha\,e^{-\beta}}{1+p^2}\Big[-p\,q+\beta(1+p^2)\arctan(p)\Big]\,.
    \label{wa-weak}
\end{align}
We fit the DE EoS, $w_\phi(a)$, obtained from our model to the equation $\hat{w}(a)=w_0+(1-a)w_a$ and determine the values of $w_0$ and $w_a$ of each model data point. The data point whose fit minimizes the $\chi^2$ value
\begin{equation}
    \chi^2=\sum_i\left(\frac{w_i(a)-\hat{w}_i(a)}{\sigma_i}\right)^2
\end{equation}
is retained. We give in Fig.~\ref{fig5} a scatter plot in the $w_0$-$w_a$ plane of the model data points which are a good fit to the DE EoS.

The left panel of Fig.~\ref{fig5} shows the Hubble parameter $H_0$ as the color axis while the right panel shows $\log\lambda$. The DESI posterior distributions from fits to $w_0w_a$CDM are shown as $1\sigma$ and $2\sigma$ ellipses corresponding to three choices of data set combinations. One can see that a large number of model data points are above the $w_0+w_a=-1$ line while fewer points are below. Also, a lot of these points fall within the $1\sigma$ and $2\sigma$ bounds from DESI.  Therefore, quintessence is able to produce values of $w_0$ and $w_a$ that are consistent with DESI without having the DE EoS cross the phantom divide. As it was argued in refs.~\cite{Wolf:2024eph,Ramadan:2024kmn,Cortes:2024lgw}, DESI's results showing an apparent phantom crossing does not mean that there is a violation of the Null Energy Condition (NEC). In fact, this phantom crossing can be a product of extrapolating the CPL parameterization of $w(a)$ outside the validity region where the distance observations are made and fitted to the model. Note that the authors of ref.~\cite{Shlivko:2024llw} have shown that some quintessence models can sometimes lie in the phantom region of the $w_0$-$w_a$ plane. The quintessence potentials adopted by the previous references differ from ours, in addition to the fact that our model involves a DM-DE interaction.

\begin{figure}[H]
\begin{centering}
\includegraphics[width=0.75\textwidth]{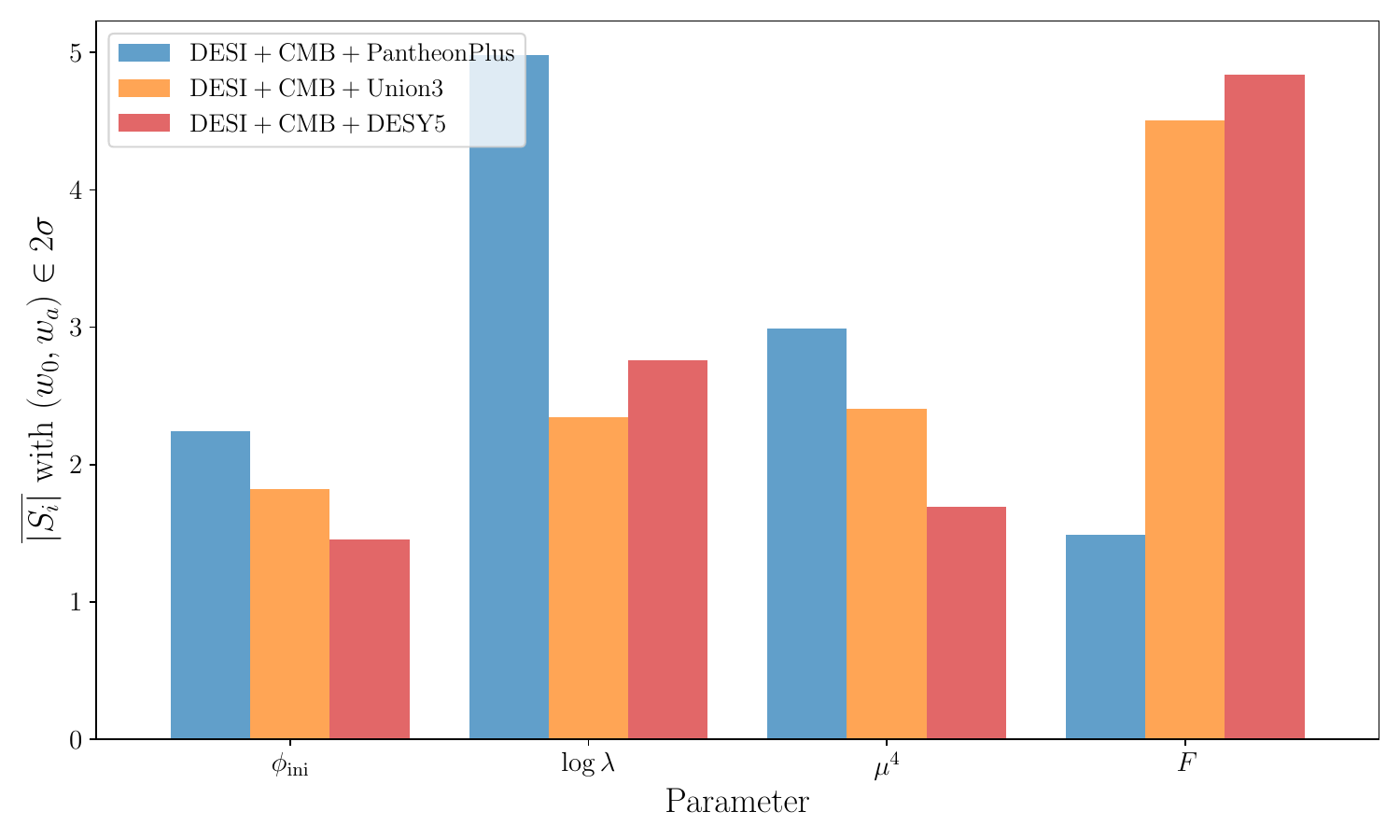}
\caption{The mean absolute Shapley value, $\overline{|S_i|}$, of the four model input parameters for the three data sets considered by DESI. The parameter with the largest $\overline{|S_i|}$ value has the most impact in bringing the $(w_0,w_a)$ values to within $2\sigma$ of their experimental value.  }
\label{fig6}
\end{centering}
\end{figure}

Our model contains four input parameters and to determine which parameter has the strongest impact in pushing the $w_0$ and $w_a$ values to within $2\sigma$ of their experimental values determined by DESI, we use \code{SHAP} (SHapley Additive exPlanations)~\cite{Lundberg:2017uca}. We use the machine learning algorithm \code{XGBoost}~\cite{Chen:2016btl} to train a gradient-boosted decision tree classifier (\code{XGBClassifier}) on the Monte Carlo data from our model. The algorithm learns how the model input parameters work together to produce points that fall within $2\sigma$ of DESI's $w_0$ and $w_a$ values. \code{XGBoost} builds a model composed of several decision trees equipped with a probability that the classifier has either rendered a point within the experimental limits or not. This information is used by \code{TreeSHAP} (part of \code{SHAP}), which is a fast algorithm used to compute the exact Shapley values. Each model input parameter is considered a `feature' and the \code{SHAP} value can tell us which particular feature in the model has the largest impact on the model prediction or outcome. Let $f(x)$ be the model prediction for input $x$, $A$ is the full set of features $x_i$, $B\subseteq A$ is the set excluding feature $x_i$ and $|A|$ ($|B|$) is the number of features in set $A$ ($B$). The \code{SHAP} value for feature $x_i$ is then
\begin{equation}
    S_i(f,x)=\sum_{B\subseteq A\backslash\{i\}}\frac{|B|!(|A|-|B|-1)!}{|A|!}\Big[f_{B\cup\{i\}}(x)-f_B(x)\Big],
\end{equation}
where $f_B(x)$ is the model prediction with feature $x_i$ withheld and $f_{B\cup\{i\}}(x)$ is the model prediction with the feature included. The mean Shapley values, $\overline{|S_i|}$, defined as
\begin{equation}
  \overline{|S_i|}=\frac{1}{N}\sum_{j=1}^N |S_i(f,x_j)|,  
\end{equation}
where the sum runs over all data points in the sample, is shown in Fig.~\ref{fig6} for the four input parameters: $\phi_{\rm ini}$, $\log\lambda$, $\mu^4$ and $F$. The analysis is done for DESI's three data sets that include PantheonPlus, Union3 and DESY5 and are shown in blue, orange and red, respectively. One can see that for the PantheonPlus data set, the DM-DE interaction strength, $\lambda$, has the largest Shapley value which means that this parameter has the largest impact in driving the $w_0$ and $w_a$ values to within $2\sigma$ of their experimental values. For the data sets, Union3 and DESY5, the parameter $F$ has the largest impact.

\begin{figure}[H]
\begin{centering}
\includegraphics[width=0.495\textwidth]{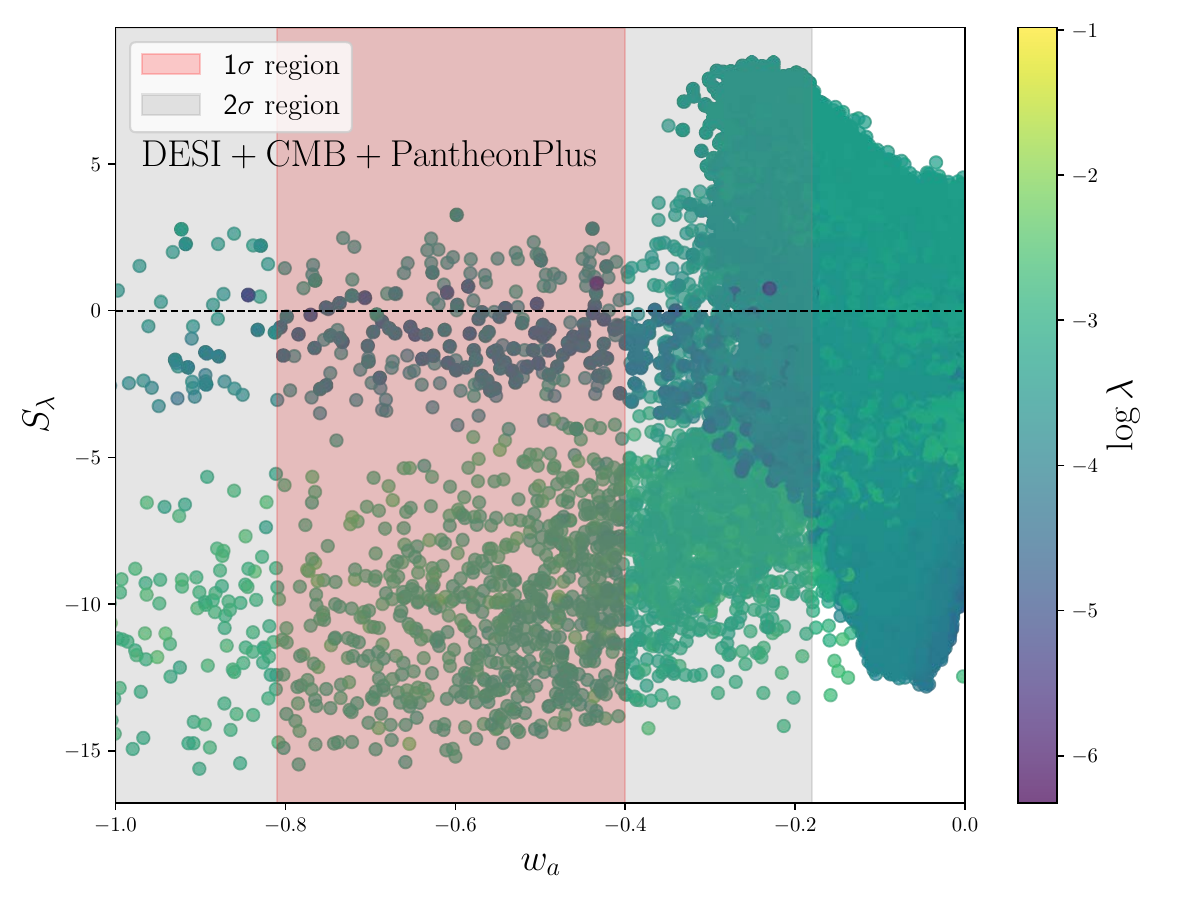}
\includegraphics[width=0.495\textwidth]{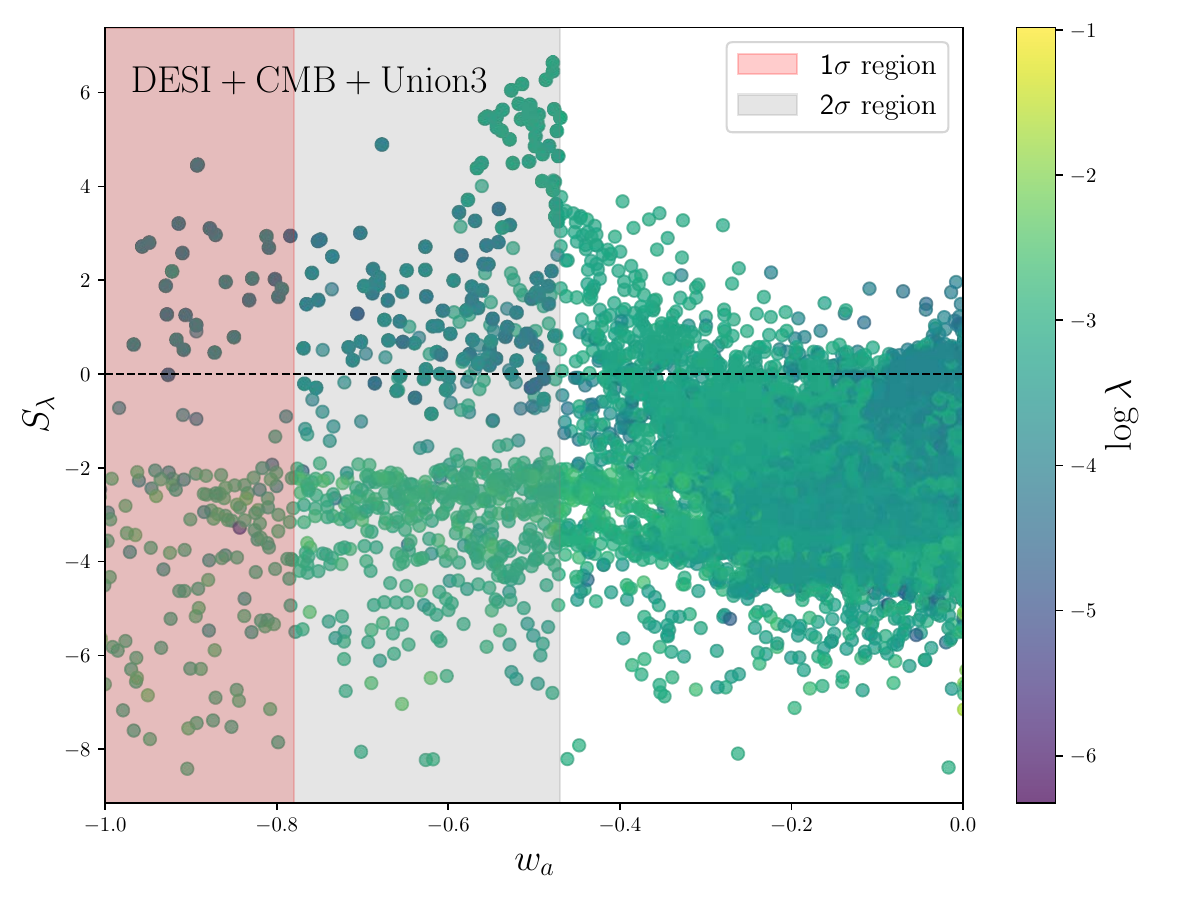}
\caption{A scatter plot of the model data points showing the Shapley value, $S_\lambda$, of the DM-DE coupling $\lambda$ versus the parameter $w_a$. The color axis represents the values of $\log\lambda$. The analysis in the left panel is done for the DESI+CMB+PantheonPlus data set while the right panel is for the DESI+CMB+Union3 data set. }
\label{fig7}
\end{centering}
\end{figure}

In the scatter plots of Fig.~\ref{fig7}, we show the data points in $S_\lambda$-$w_a$ plane (with $S_\lambda$ the Shapley value of $\lambda$) and the color axis represents the DM-DE interaction strength. The left panel shows the PantheonPlus data set while the right one shows Union3. One can see that for the PantheonPlus data set, the highest density of points for higher $\lambda$ occurs above $S_\lambda=0$ and are concentrated near the upper edge of the $2\sigma$ region which confirms that $\lambda$ is an impactful parameter in this data set. However, for the Union3 data set (right panel), most points are below $S_\lambda=0$ and the few that have $S_\lambda>0$ are characterized by small values of $\lambda$.  

We give in Table~\ref{tab1} the 68\% CL values of the model input parameters and some derived cosmological parameters. The obtained values of $w_0$, $w_a$, $H_0$ and $\Omega_{\rm m}$ agree with DESI's results for each of the three data sets. Compared to the other data sets, the PantheonPlus data set gives the largest upper limit on $\lambda$ value. We show in the last row the difference in the Deviance Information Criterion (DIC)~\cite{Spiegelhalter:2002yvw,Trotta:2008qt} defined as
\begin{equation}
    \Delta(\mathrm{DIC})=\mathrm{DIC}_{\lambda=0}-\mathrm{DIC}_{\lambda\neq 0}
\end{equation}
One can see that for all three data sets, $\Delta(\mathrm{DIC})<0$ which means that the no interaction model is strongly preferred over the DM-DE interaction scenario.

\begin{table}[H]
\centering
{\tabulinesep=1.2mm
\begin{tabu}{cccc}
\hline\hline
Parameter & DESI+CMB+PantheonPlus & DESI+CMB+Union3 & DESI+CMB+DESY5 \\
\hline
$F$ [$m_{\rm Pl}$] & $1.14^{+1.28}_{-0.57}$ & $1.22^{+1.24}_{-0.96}$ & $0.336^{+1.082}_{-0.243}$  \\
$\phi_{\rm ini}$ [$m_{\rm Pl}$] & $1.11^{+0.22}_{-0.19}$ & $1.10^{+0.66}_{-2.22}$ & $1.97^{+0.50}_{-1.34}$ \\
$10^8\mu^4$ & $2.86^{+0.61}_{-0.65}$ & $2.38^{+1.29}_{-0.39}$ & $2.95^{+1.87}_{-1.30}$ \\
$\log\lambda$ & $<-2.78$ & $<-2.99$ & $<-3.26$ \\ 
$w_0$ & $-0.80^{+0.73}_{-0.03}$ & $-0.518^{+0.180}_{-0.321}$ & $-0.813^{+0.353}_{-0.031}$  \\ 
$w_a$ & $-0.227^{+0.044}_{-0.467}$ & $-0.505^{+0.282}_{-0.388}$ & $-0.556\pm 0.366$  \\
$H_0$ & $67.97_{-2.80}^{+5.83}$ & $68.17_{-3.10}^{+5.63}$ & $70.59_{-4.74}^{+3.25}$  \\
$\Omega_{\rm m}$ & $0.308_{-0.047}^{+0.027}$ & $0.306_{-0.045}^{+0.030}$ & $0.286_{-0.025}^{+0.043}$  \\
$\Delta(\mathrm{DIC})$ & $-109$ & $-29.6$ & $-71.5$  \\
\hline\hline
\end{tabu}}
\caption{Constraints on some of the cosmological parameters of our model. The values are quoted at 68\% CL intervals, unless an upper or lower bounds are shown, in which case it is the 95\% CL interval.}
\label{tab1}
\end{table}

\begin{figure}[H]
\begin{centering}
\includegraphics[width=0.495\textwidth]{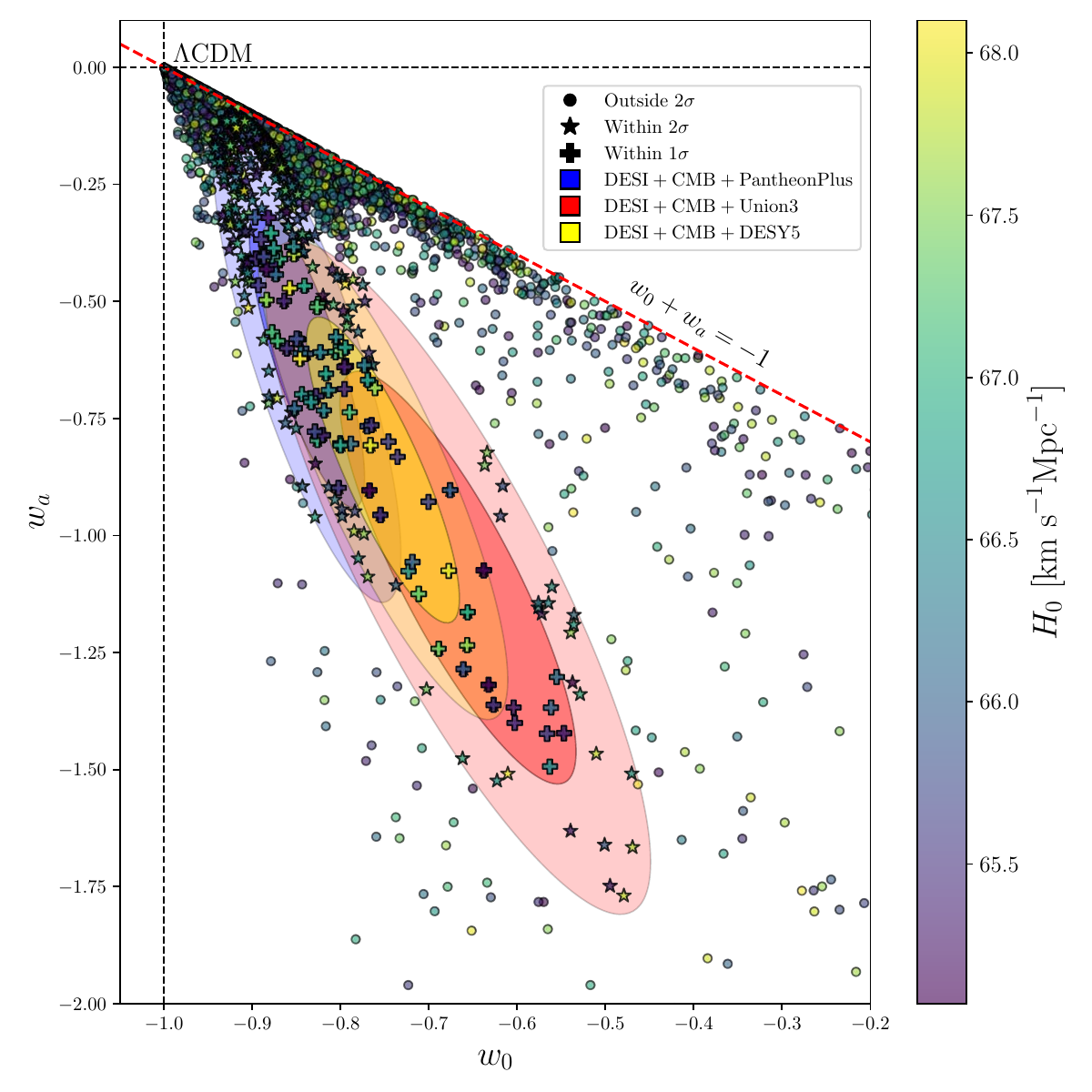}
\includegraphics[width=0.495\textwidth]{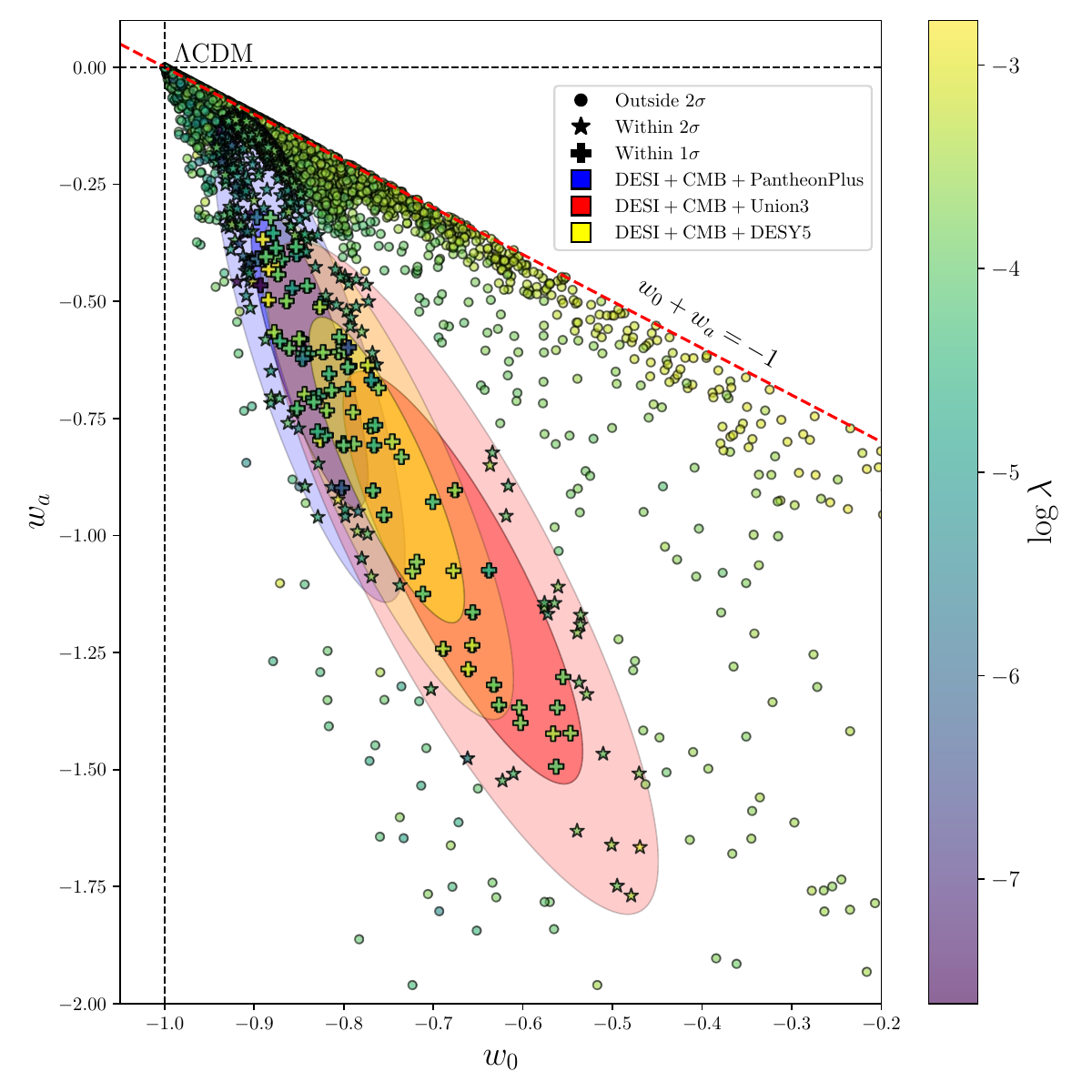}
\caption{Scatter plots in the $w_0$-$w_a$ plane similar to Fig.~\ref{fig5} but now with the upper limit on $\log\lambda$ imposed. To enforce matter domination at the right epoch, only points below $w_0+w_a=-1$ are considered. Many of the remaining points still fall within the $1\sigma$ and $2\sigma$ experimental constraints.  }
\label{fig8}
\end{centering}
\end{figure}

With the DESI DR2 results, we are able to set an upper limit on the DM-DE interaction strength. We show in Fig.~\ref{fig8} the model points that respect the imposed upper limit on $\lambda$, as well as constraints on $H_0$ and $\Omega_{\rm m}$. Many points still populate the $1\sigma$ and $2\sigma$ posterior distributions in the $w_0$-$w_a$ plane even for $\lambda$ values close to the upper limit.

A final word regarding the calculation of $\Delta$(DIC) is warranted. The DIC is determined as $-2\langle\log\mathcal{L}\rangle+p_D$, where the first term is the mean deviance and the second term is a penalty on model complexity. Our interacting model has an extra parameter $\lambda$ over the non-interacting model which gets penalized leading to $\Delta(\mathrm{DIC})<0$. A future direction for this work would be to investigate field-theoretic interaction terms than can overcome this difficulty, making a DM-DE interaction model more favorable.

\section{Extraction of cosmological parameters with Bayesian inference}

Dark matter-dark energy interaction will also impact density perturbations and leave noticeable effects on the temperature and matter power spectra. So, in order to extract realistic constraints on the cosmological parameters, we confront the model with the most recent cosmological data, namely, CMB measurements from Planck 2018, baryon acoustic oscillations from DESI-DR2 as well as a number of local measurements from supernova data.  

We solve, in \code{CLASS}, the KG perturbation equations (in the synchronous gauge) of the fields $\phi$ and $\chi$, Eqs.~(\ref{kgp1}) and~(\ref{kgc1}), along with their background equations.  The background and perturbation equations of all the other species are evolved as in $\Lambda$CDM. We set as initial conditions, $\chi_1=\chi^\prime_1=0$ and $\phi_1=\phi^\prime_1=0$. This is analogous to setting $\delta_{\rm ini}=0$ and $\Theta_{\rm ini}=0$. Despite being set to zero initially, these quantities are quickly driven to the attractor solution~\cite{Ballesteros:2010ks}. A remark regarding the synchronous gauge is in order. This choice is known to not completely fix the gauge degrees of freedom and in $\Lambda$CDM we rely on the fact that $w_{\rm CDM}=0$ to fix the gauge. For a scalar field, this is not the case throughout its evolution as $w_\chi$ is dynamical. In order to be able to still consider the synchronous gauge in \code{CLASS}, we allow for a small amount of CDM by setting $\Omega_{\rm CDM} h^2=10^{-5}$. 

In our analysis, we use the following data sets:
\begin{enumerate}
\item The Planck 2018 temperature anisotropies and polarization measurements. The temperature and polarization (TT TE EE) likelihoods include low multipole data ($\ell<30$)~\cite{Planck:2018vyg,Planck:2018nkj,Planck:2019nip}. The high multipole likelihood includes: $30\lesssim \ell\lesssim 2500$ for the TT spectrum and $30\lesssim \ell\lesssim 2000$ for the TE and EE spectra. The low-E polarization likelihood includes $2\leq \ell\leq 30$ for the EE spectrum.  We also include the Planck 2018 lensing likelihood~\cite{Planck:2018lbu} which is inferred from the lensing potential power spectrum. This whole data set is referred to as \textbf{CMB}.
\item Baryon Acoustic Oscillations (from DESI-DR2): We include BAO measurements from DESI's second data release, which takes into account observations of galaxies and quasars~\cite{DESI:2025zgx}, as well as Lyman-$\alpha$ tracers~\cite{DESI:2025zpo}. These measurements cover both isotropic and anisotropic BAO constraints over $0.295\leq z\leq 2.330$, divided into nine redshift bins. This data set is referred to as \textbf{DESI}.
\item The combination PantheonPlus+SH0ES~\cite{Brout:2022vxf,Riess:2021jrx} data set which uses an additional Cepheid distance as a calibrator of the Supernova SNIa intrinsic magnitude. This data set is referred to as \textbf{PPS}. 
\item Union 3.0: The Union 3.0 database, consisting of 2087 SN Ia within the range $0.001<z<2.260$~\cite{Rubin:2023jdq}. The data set has 1363 SN Ia in common with the PantheonPlus sample and uses a special and different treatment of systematic errors and uncertainties by employing Bayesian hierarchical modeling. We refer to this dataset as \textbf{Union3}.
\item DESY5: In their Year 5 data release, the Dark Energy Survey (DES) presented results based on a newly published, uniformly selected sample of 1635 photometrically classified Type Ia supernovae, covering redshifts from $0.1 < z < 1.3$~\cite{DES:2024jxu}. This sample is supplemented by 194 low-redshift SN Ia, drawn from the PantheonPlus dataset, within the range $0.025 < z < 0.1$. We refer to this combined dataset as \textbf{DESY5}.
\end{enumerate}

The Boltzmann solver \code{CLASS} is interfaced with \code{Cobaya}~\cite{Torrado:2020dgo}, a code for sampling and statistical modeling, to perform a Markov Chain Monte Carlo (MCMC) analysis of the model parameter space which allows us to extract constraints on the cosmological parameters through Bayesian inference. \code{Cobaya} utilizes an adaptive, speed-hierarchy-aware MCMC sampler (adapted from \code{CosmoMC})~\cite{Lewis:2002ah,Lewis:2013hha} using the fast-dragging procedure described in ref.~\cite{Neal:2005uqf}. We monitor the convergence of the chains using the Gelman-Rubin~\cite{Gelman:1992zz} criterion $R-1<0.05$. After convergence, the chains are analyzed with \code{GetDist}~\cite{Lewis:2019xzd}\footnote{\url{https://github.com/cmbant/getdist}}, a package allowing for the extraction of numerical results, including 1D posteriors and 2D marginalized probability contours.

The sampling parameters consist of the baseline $\Lambda$CDM parameters along with the four additional free parameters of our model
\begin{equation}
    \underbrace{\Omega_{\rm b} h^2, ~~\Omega_\chi, ~~z_{\rm reio}, ~~100\theta_s, ~~\log(10^{10}A_s), ~~n_s}_{\Lambda\text{CDM}}, ~~\log\mu^4,~~F,~~\phi_{\rm ini},~~\log\lambda\,,
\end{equation}
where $\Omega_\chi$ ($\Omega_{\rm b}$) is the DM (baryon) fractional density, $A_s$ is the amplitude of primordial fluctuations, $n_s$ is the spectral index of the primordial power spectrum, $\theta_s$ is the angular scale of the sound horizon at the surface of last scattering, and $z_{\rm reio}$ is the redshift at reionization. We impose flat priors on all the parameters.

We begin our discussion by examining the findings of the analysis after including perturbations. They are summarized in Table~\ref{tab2}. The table shows the 68\% CL constraints on the cosmological parameters under the four data set combinations mentioned above. The upper set of parameters in the table are the ones being sampled and the lower set (separated by the double line) are the derived parameters. The sampled parameters are driven to the values shown in the table that best fit the data. The baryon density fraction remains almost stable across all data sets while the DM density fraction shows more than a 5\% increase in the DESY5 and Union3 data sets as compared to the first two sets. The model-specific $\mu^4$ parameter has values ranging from $\sim 5.5\times 10^{-8}~\mathrm{m}_{\rm Pl}^2/\text{Mpc}^2$ (CMB+DESI) to $\sim 7.6\times 10^{-8}~\mathrm{m}_{\rm Pl}^2/\text{Mpc}^2$ (CMB+DESI+Union3). The strongest lower bound imposed on the parameter $F$ comes from the CMB+DESI+PPS data set, with $F>0.62$. The lower bound relaxes to $F\sim 0.29$ with the inclusion of DESY5. Furthermore, we are able to set an upper limit on $\phi_{\rm ini}$, with the PPS data set imposing the stringent constraint. All three data sets indicate a weak DM-DE coupling with $\lambda\lesssim 10^{-3}$ being the weakest upper limit and $\lambda\lesssim 10^{-5.7}$ being the strongest. These limits are stronger than the ones derived used only the background evolution. Addressing the Hubble tension, the model can only moderately alleviate it at the 95\% CL with the highest value being driven by the PantheonPlus+SH0ES data set. The root-mean-square mass fluctuation amplitude for spheres of size $8h^{-1}$Mpc, $\sigma_8$, is another important parameter that can be constrained by our model. A related parameter $S_8\equiv \sigma_8\sqrt{\Omega_{\rm m}/0.3}$ is usually employed as a degeneracy-breaking parameter and to express the tension in measurements of $\sigma_8$ for weak lensing and CMB probes. The values of $S_8$ we obtain are consistent with the results from the recent \code{KiDS-Legacy} survey~\cite{Wright:2025xka} analysis.

\begin{table}[H]
\centering
{\tabulinesep=1.2mm
\resizebox{\textwidth}{!}{\begin{tabu}{ccccc}
\hline\hline
\textbf{Parameter} & \textbf{CMB+DESI} & \textbf{CMB+DESI+PPS} & \textbf{CMB+DESI+DESY5} & \textbf{CMB+DESI+Union3} \\
\hline
{$\log(10^{10} A_\mathrm{s})$} & $3.042\pm 0.010            $ & $3.046\pm 0.013            $ & $3.044\pm 0.011            $ & $3.045\pm 0.012            $\\
{$n_\mathrm{s}   $} & $0.9684\pm 0.0036          $ & $0.9703\pm 0.0033          $ & $0.9696\pm 0.0037          $ & $0.9694\pm 0.0035          $\\
{$100\theta_\mathrm{s}$} & $1.04217\pm 0.00029        $ & $1.04227\pm 0.00027        $ & $1.04219\pm 0.00028        $ & $1.04220\pm 0.00028        $\\
{$\Omega_\mathrm{b} h^2$} & $0.02278\pm 0.00011        $ & $0.02283\pm 0.00011        $ & $0.02278\pm 0.00011        $ & $0.02279\pm 0.00011        $\\
{$\Omega_\chi    $} & $0.2420\pm 0.0083          $ & $0.2415\pm 0.0066          $ & $0.256^{+0.011}_{-0.0097}  $ & $0.2543^{+0.0093}_{-0.012} $\\
{$z_\mathrm{reio}$} & $7.69\pm 0.48              $ & $7.86\pm 0.61              $ & $7.80\pm 0.56              $ & $7.84\pm 0.60              $\\
{$\log\mu^4      $} & $-7.233^{+0.013}_{-0.028}  $ & $-7.230^{+0.013}_{-0.024}  $ & $-7.192^{+0.041}_{-0.065}  $ & $-7.175^{+0.053}_{-0.075}  $\\
{$F~[m_{\rm Pl}]$} & $> 0.561                   $ & $> 0.620                   $ & $0.51^{+0.20}_{-0.22}      $ & $0.63^{+0.23}_{-0.17}      $\\
{$\phi_{\mathrm{ini}}~[m_{\rm Pl}]$} & $< 0.157                   $ & $< 0.263                   $ & $< 0.384                   $ & $< 0.560                   $\\
{$\log\lambda    $} & $-4.9^{+2.0}_{-2.7}        $ & $-5.49^{+0.97}_{-2.2}      $ & $< -5.49                   $ & $< -5.69                   $\\
\hline\hline
$H_0~[\rm km/s/Mpc]$ & $69.4^{+1.1}_{-0.98}       $ & $69.39\pm 0.87             $ & $67.6^{+1.3}_{-1.6}        $ & $67.8^{+1.6}_{-1.3}        $\\
$\Omega_\mathrm{m}         $ & $0.276^{+0.029}_{-0.016}   $ & $0.280^{+0.025}_{-0.011}   $ & $0.294^{+0.028}_{-0.015}   $ & $0.295^{+0.024}_{-0.015}   $\\
$\Omega_\phi               $ & $0.7092\pm 0.0097          $ & $0.7097^{+0.0079}_{-0.0071}$ & $0.693\pm 0.012            $ & $0.695^{+0.014}_{-0.011}   $\\
$\tau_\mathrm{reio}          $ & $0.0559\pm 0.0049          $ & $0.0579^{+0.0060}_{-0.0068}$ & $0.0571^{+0.0054}_{-0.0061}$ & $0.0575\pm 0.0062          $\\
$S_8                       $ & $0.769^{+0.034}_{-0.018}   $ & $0.773^{+0.028}_{-0.014}   $ & $0.778^{+0.027}_{-0.014}   $ & $0.780^{+0.023}_{-0.012}   $\\
$w_0                       $ & $-0.900^{+0.069}_{-0.25}   $ & $-0.937^{+0.021}_{-0.17}   $ & $-0.925^{+0.055}_{-0.13}   $ & $-0.890^{+0.075}_{-0.16}   $\\
$w_a                       $ & $-0.018^{+0.035}_{-0.016}  $ & $-0.018^{+0.029}_{-0.013}  $ & $-0.076^{+0.11}_{-0.046}   $ & $-0.109^{+0.14}_{-0.070}   $\\
\hline
$\Delta\chi^2_{\rm min}$ & 25.11 & 21.61 & 14.23 & 21.05 \\
$\Delta\mathrm{AIC}$ & 33.11 & 29.61 & 22.23 & 29.05 \\
\hline\hline
\end{tabu}}}
\caption{Constraints on some of the cosmological parameters of our model. The values are quoted at 68\% CL intervals for three data set combinations. The middle double line separates the sampled and derived parameters using MCMC. In the last two rows we show the values of $\Delta\text{AIC}\equiv\text{AIC}_{\rm QCDM}-\text{AIC}_{\Lambda\rm CDM}$ and $\Delta\chi^2_{\rm min}\equiv\chi^2_{\rm QCDM,min}-\chi^2_{\rm \Lambda CDM,min}$.  }
\label{tab2}
\end{table}

The values of the parameters $w_0$ and $w_a$ based on Eqs.~(\ref{w0-weak}) and~(\ref{wa-weak}) and obtained with the inclusion of the latest DESI-DR2 results show values consistent with $\Lambda$CDM, i.e., $w_0\simeq -1$ and $w_a=0$. But the central values of $w_0$ and $w_a$ (and within the $1\sigma$ and $2\sigma$ limits) indicate an evolving DE equation of state. We see from table~\ref{tab2} that the inclusion of the supernova data sets greatly impacts the values of $w_0$ and $w_a$, especially the presence of DESY5 and Union3. We will have more to say about $w_0$ and $w_a$ later.

\begin{figure}[H]
\begin{centering}
\includegraphics[width=1.0\textwidth]{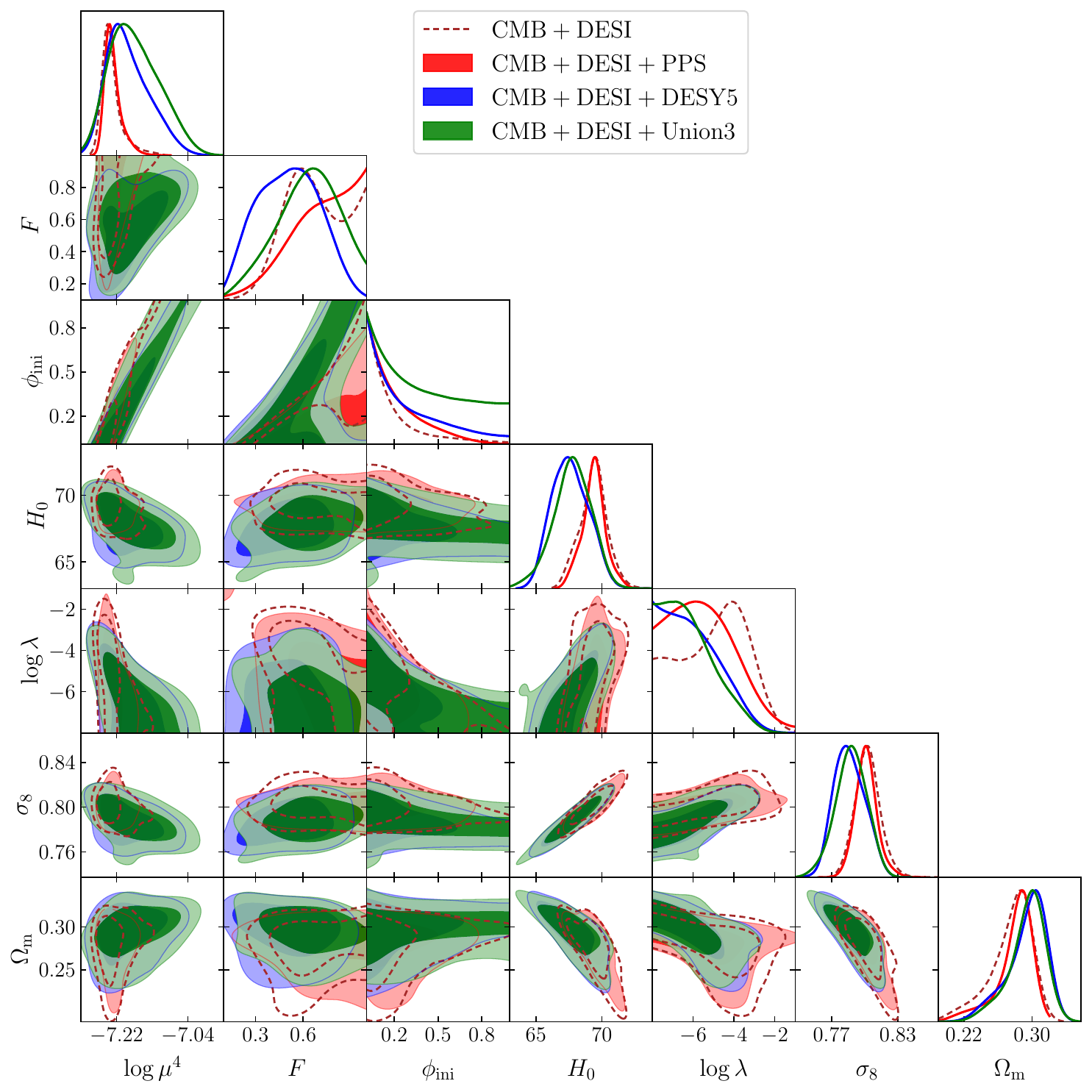}
\caption{Triangle plot showing the 2D joint and 1D marginalized posteriors of $\log\mu^4$, $\phi_{\rm ini}$, $F$, $\log\lambda$, $H_0$, $\sigma_8$ and $\Omega_{\rm m}$ in the interacting dark matter-dark energy model for the following data set combinations: CMB+DESI+PPS (red contours), CMB+DESI+DESY5 (blue contours) and CMB+DESI+Union3 (green contours). }
\label{fig9}
\end{centering}
\end{figure}

To evaluate how good the interacting model, QCDM, fit the data in comparison to $\Lambda$CDM, we use the Akaike Information Criterion (AIC)~\cite{Akaike:1974vps} which is defined as
\begin{equation}
    \text{AIC}\equiv -2\ln\mathcal{L}_{\rm max}+2K,
\end{equation}
where $\mathcal{L}_{\rm max}$ is the maximum likelihood of the model, and $K$ is the number of free parameters. The difference in AIC is defined as $\Delta\text{AIC}\equiv\text{AIC}_{\rm QCDM}-\text{AIC}_{\Lambda\rm CDM}$. As a rule of thumb, $\Delta\text{AIC}<-5$ indicates a strong preference for QCDM over $\Lambda$CDM, while $\Delta\text{AIC}>10$ indicates a decisive preference for $\Lambda$CDM. The last row in table~\ref{tab2} shows the latter case, i.e., the data prefers $\Lambda$CDM over QCDM for all data set combinations. 

The 2D marginalized posteriors of a selection of parameters are shown in the triangle plot of Fig.~\ref{fig9} for the three data set combinations: $\mathrm{CMB+DESI}$ (brown dashed contour), $\mathrm{CMB+DESI+PPS}$ (red contours), $\mathrm{CMB+DESI+DESY5}$ (blue contours) and $\mathrm{CMB+DESI+Union3}$ (green contours). Some interesting positive correlations can be seen between the theory parameters, $\log\mu^4$, $F$ and $\phi_{\rm ini}$. A moderate correlation is seen between $H_0$ and $\log\lambda$ and between $\sigma_8$ and $\log\lambda$. Notice how the presence of the SN Ia data sets shift the constraints on the different parameters, for example, setting tighter constraints on $\log\lambda$ while relaxing constraints on $\phi_{\rm ini}$. The DESI DR2 results show a shift in the measurement of the matter density $\Omega_{\rm m}$ to a higher value with uncertainties reduced by around a factor of two. From Fig.~\ref{fig9}, there is a negative correlation between $\Omega_{\rm m}$ and $\log\lambda$, meaning that stronger DM-DE interaction pushes $\Omega_{\rm m}$ to lower values. Since the DESI DR2 measurement indicate higher $\Omega_{\rm m}$, then $\log\lambda$ has become more constrained towards a weaker coupling value.     

\begin{figure}[H]
\begin{centering}
\includegraphics[width=0.495\textwidth]{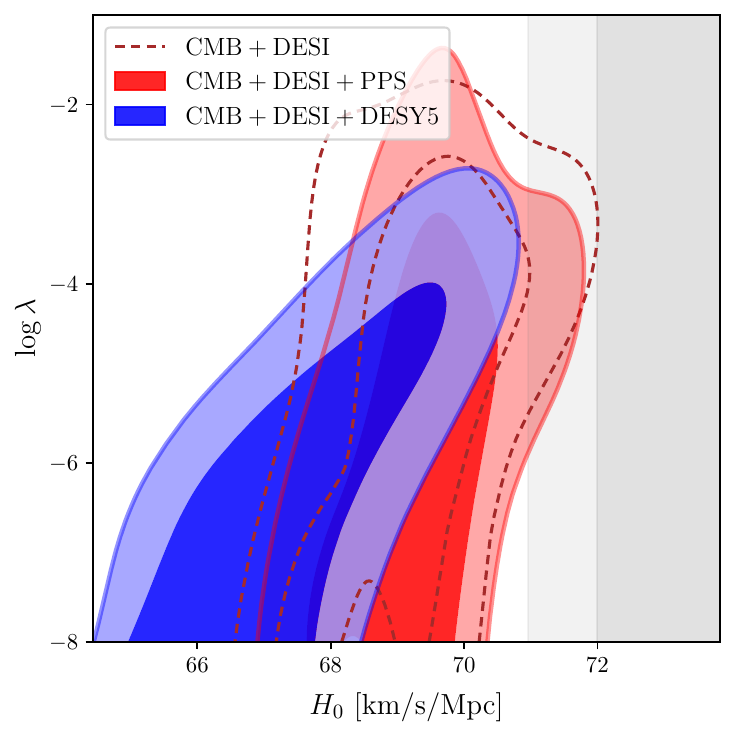}
\includegraphics[width=0.495\textwidth]{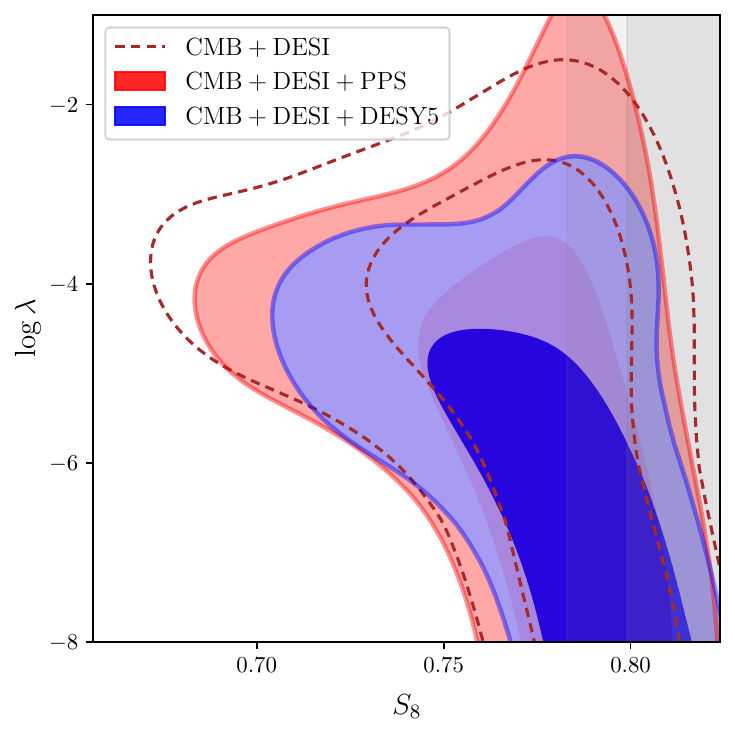}
\caption{The 2D contours at 68\% and 95\% CL for the DM-DE coupling strength and $H_0$ (left panel) and $S_8$ (right panel). The gray bands in the left panel correspond to the SH0ES~\cite{Riess:2021jrx} value $H_0=73.04+\pm 1.04$ km/s/Mpc and the ones in the right panel correspond to the \code{KiDS}-Legacy~\cite{Wright:2025xka} result of $S_8=0.815^{+0.016}_{-0.021}$.   }
\label{fig10}
\end{centering}
\end{figure}

The contours of Fig.~\ref{fig10} show the correlation between $H_0$ and $\log\lambda$ (left panel) and $S_8$ and $\log\lambda$ (right panel). The gray bands show the $1\sigma$ and $2\sigma$ corridors for the experimental values of $H_0$ and $S_8$. The $1\sigma$ contours of $H_0$ for the two data sets do not coincide with the experimental value of $H_0$, while those of $S_8$ do. The tension in $H_0$ is thus mildly alleviated in this model, while no significant tension exists in $S_8$.

\begin{figure}[H]
\begin{centering}
\includegraphics[width=0.49\textwidth]{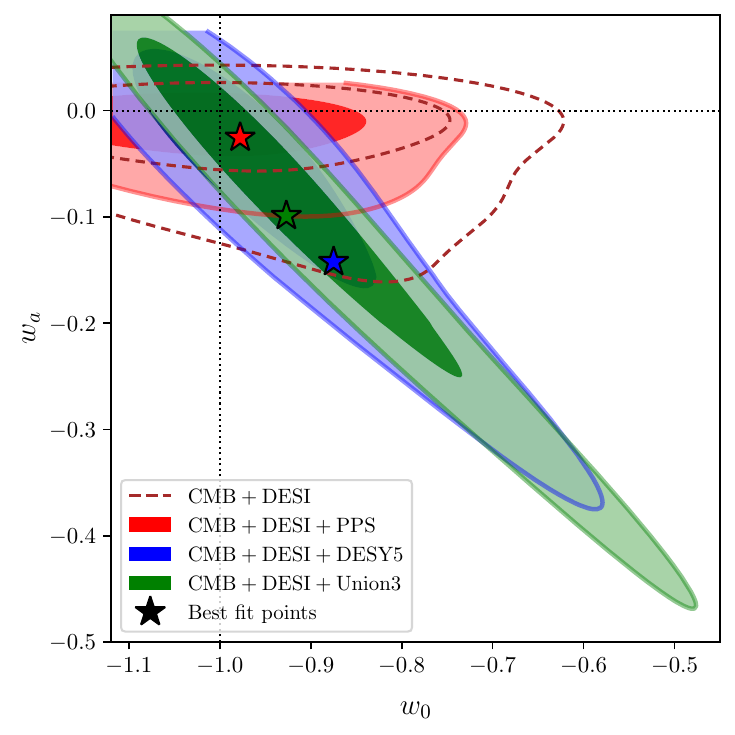}
\includegraphics[width=0.49\textwidth]{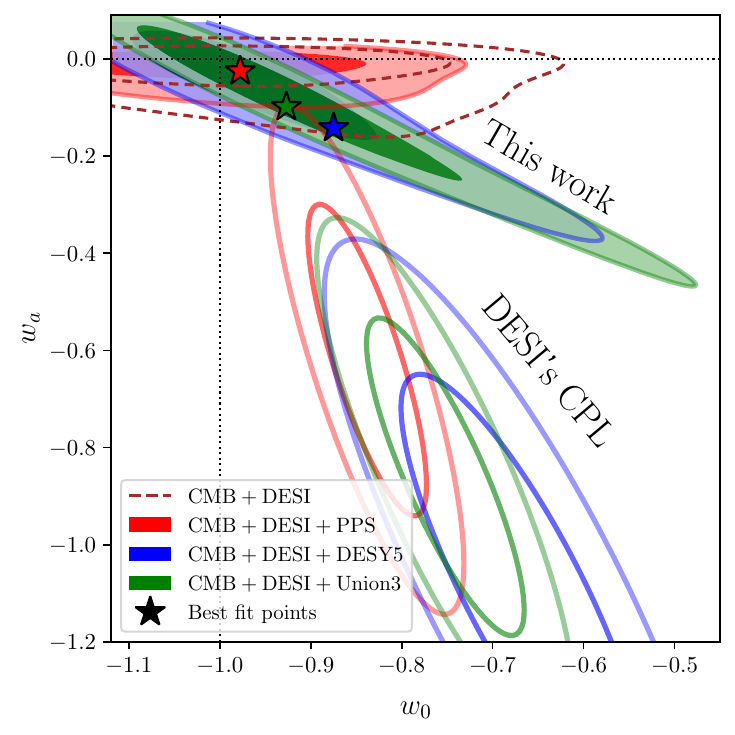}
\caption{Left panel: Results for the posterior distributions of $w_0$ and $w_a$ for the four data set combinations: CMB+DESI (dashed contours), CMB+DESI+PPS (red contours), CMB+DESI+DESY5 (blue contours) and CMB+DESI+Union3 (green contours). The intersection of the dotted vertical and horizontal lines corresponds to $\Lambda$CDM ($w_0=-1$ and $w_a=0$). The stars represent the best fit points for the PPS, DESY5 and Union3 data sets. Right panel: same as left panel with the addition of DESI's (unfilled) contours. }
\label{fig11}
\end{centering}
\end{figure}

Finally, we discuss the effect of the different data sets on the values of $w_0$ and $w_a$ and implications on evolving dark energy. To this end, we show in the left panel of Fig.~\ref{fig11} the $1\sigma$ and $2\sigma$ contours in the $w_0$-$w_a$ plane for the four data set combinations. Large parts of the contours lie in the fourth quadrants where $w_0<0$ and $w_a<0$, but, unlike the DESI-DR2 results, some parts extend to the other quadrants and overlapping with the point where $\Lambda$CDM lives, $(w_0,w_a)=(-1,0)$. As we have already mentioned, the values are consistent with $w_0=-1$ but the $1\sigma$ and $2\sigma$ values leave a lot of room for an evolving dark energy near $a=1$. The best fit points shown as stars for each data set (PPS, DESY5, Union3) all lie in the fourth quadrant, thus favoring evolving DE. Unlike the PantheonPlus+SH0ES, the inclusion of the DESY5 and Union3 data sets drives the $w_0$ and $w_a$ values toward an evolving DE scenario while remaining in the quintessence regime (no crossing of the phantom divide). We should note here is that even though we are using a CPL parametrization of the DE EoS, $w_0$ and $w_a$ are not free parameters. They are derived parameters and depend on the underlying particle physics model. To show the difference between this work and DESI's analysis, we add DESI's (unfilled) contours in the right panel of Fig.~\ref{fig11}. These contours are deeper in the fourth quadrant compared to the contours of this work, which is a reflection of the phenomenon of phantom divide crossing based on CPL.

\section{Conclusion}

In this work we discussed a field-theoretic model of cosmology with an interaction term between dark matter and dark energy which are described by ultralight spin zero fields. We show that the fluid formalism used to describe dark matter-dark energy interaction is inconsistent in a field-theory approach. Furthermore, considering a specific dark energy potential and a dark matter-dark energy interaction term, we study the cosmological implication of such a model on the evolution of the dark energy equation of state. We distinguish between two regimes based on the DM-DE interaction strength: the strong coupling regime with $\lambda\geq 10^{-2}$ and the weak coupling region with $\lambda<10^{-2}$. The strong coupling regime induces a transmutation of the dark energy equation of state which is visible in its evolution from a thawing behavior to a scaling freezing behavior. A dark energy equation of state from a freezing quintessence at current times is at odds with the DESI results and so the strong coupling regime can be ruled out. The weak coupling regime, on the other hand, is shown to produce results consistent with DESI, aligned with the prediction of an evolving equation of state. We provide a fit to the dark energy equation of state, which, to leading order in $(1-a)$, can produce $w_0$ and $w_a$ values consistent with DESI without crossing the phantom divide. Using the three-data set combinations from DESI, we derive an upper limit on the value of $\lambda$ and show that many benchmark points close to this upper limit remain consistent with DESI. After that, we derived more stringent constraints on the different cosmological parameters by including the effect of cosmological perturbations. The final results indicate that the DE in QCDM includes the possibility of a cosmological constant while leaving ample room for an evolving DE equation of state. This is manifest in the fact that the best fit points for the three data sets all lie in the fourth quadrant of the $w_0$-$w_a$ plane.  

\vspace{1cm}

{\bf Acknowledgments:} A communication with Kyle S. Dawson is acknowledged. 
The research of PN was supported in part by the NSF Grant PHY-2209903. The analysis presented here was done using the computing resources of the Phage Cluster at Union College.

\appendix

\section*{Appendices}

\section{On consistent fluid equations for interacting fields}\label{appA} 

In this appendix  we give further details of the fluid equations arising from a 
 Lagrangian approach which generates sources to the fluid equations
 in a consistent manner once the interaction between the fields is specified in the Lagrangian.
For simplicity we will work with cosmic time and then convert the final result for conformal time.
For two interacting scalar fields $\phi$ and $\chi$ the total Lagrangian is given by 
\begin{equation}
    \mathcal{L} = -\frac{1}{2} \partial_\mu \phi \partial^\mu \phi - V_1(\phi) - \frac{1}{2} \partial_\mu \chi \partial^\mu \chi - V_2(\chi) - V_{12}(\phi, \chi),
\end{equation}
where $V_{12}(\phi, \chi)$ specifies the interaction between the fields $\phi$ and $\chi$. 
The partial stress tensors for $\phi$ and $\chi$ are given by
\begin{align}
    T_{\mu\nu}^{(\phi)} &= \partial_\mu \phi \partial_\nu \phi - g_{\mu\nu} \left[
         \frac{1}{2} \partial_\lambda \phi \partial^\lambda \phi + V_1(\phi) + V_{12}(\phi, \chi) \right], \\
    T_{\mu\nu}^{(\chi)} &= \partial_\mu \chi \partial_\nu \chi - g_{\mu\nu} \left[ \frac{1}{2} \partial_\lambda \chi \partial^\lambda \chi + V_2(\chi) + V_{12}(\phi, \chi) \right].
\end{align}
The total stress tensor is given by
\begin{equation}
    T_{\mu\nu}^{\text{total}} = \partial_\mu \phi \partial_\nu \phi + \partial_\mu \chi \partial_\nu \chi - g_{\mu\nu} \left[ 
     \frac{1}{2} \partial_\lambda \phi \partial^\lambda \phi +  \frac{1}{2} \partial_\lambda \chi \partial^\lambda \chi  + V_1 + V_2 + V_{12} \right].
\end{equation}
Assuming spatial homogeneity (e.g., $\phi_i = \phi_i(t)$), we have:
\begin{align}
    \rho_\phi &= \frac{1}{2} \dot{\phi}^2 + V_1(\phi) + V_{12}(\phi, \chi), \\
    p_\phi &= \frac{1}{2} \dot{\phi}^2 - V_1(\phi) - V_{12}(\phi, \chi), \\
    \rho_\chi &= \frac{1}{2} \dot{\chi}^2 + V_2(\chi) + V_{12}(\phi, \chi), \\
    p_\chi &= \frac{1}{2} \dot{\chi}^2 - V_2(\chi) - V_{12}(\phi, \chi).
\end{align}
The above lead to the total energy density and pressure so that
\begin{align}
  \rho\equiv  \rho_{\text{total}} &= \frac{1}{2} (\dot{\phi}^2 + \dot{\chi}^2) + V_1 + V_2 + V_{12}, 
  \\
  p\equiv  p_{\text{total}} &= \frac{1}{2} (\dot{\phi}^2 + \dot{\chi}^2) - V_1 - V_2 - V_{12}.
\end{align}
One may note the following relation between $\rho$ and ($\rho_\phi+
\rho_\chi$) and a similar relation between $p$ and ($p_\phi+p_\chi$).
  \begin{align}
  \label{appen1}
 \rho&= \rho_\phi+ \rho_\chi -V_{12}, \\
  p&= p_\phi+ p_\chi +V_{12}.
  \label{appen2}
 \end{align}
 Next we compute the fluid equations for the fields $\phi$ and $\chi$. Here we begin with the
 Klein-Gordon equations in an expanding universe assuming spatial homogeneity, so that  
the equations $\phi$ and $\chi$ are
\begin{align}
    \ddot{\phi} + 3H\dot{\phi} + \frac{\partial V_1}{\partial \phi} + \frac{\partial V_{12}}{\partial \phi} &= 0, \\
    \ddot{\chi} + 3H\dot{\chi} + \frac{\partial V_2}{\partial \chi} + \frac{\partial V_{12}}{\partial \chi} &= 0.
\end{align}
Using the equations of motion the corresponding individual continuity equations for the densities for $\phi$ and $\chi$ directly follow and are given by
\begin{align}
\label{appen3}
    \dot{\rho}_\phi + 3H(\rho_\phi + p_\phi) &= \dot{\chi} \frac{\partial V_{12}}{\partial \chi} 
    \equiv Q_\phi\,, \\
    \dot{\rho}_\chi + 3H(\rho_\chi + p_\chi) &= \dot{\phi} \frac{\partial V_{12}}{\partial \phi}
    \equiv Q_\chi\,.
    \label{appen4}
\end{align}
Note that 
\begin{align}
Q_\phi+Q_\chi&=  \dot{\chi} \frac{\partial V_{12}}{\partial \chi} +  \dot{\phi} \frac{\partial V_{12}}{\partial \phi}=\dot {V}_{12}\,.
\label{appen5}
\end{align}
Using Eqs.~(\ref{appen1}),~(\ref{appen2}) and Eqs.~(\ref{appen3})$-$(\ref{appen5})
one finds that the total density $\rho$ satisfies the relation
\begin{equation}
    \dot{\rho} + 3H(\rho + p) = 0.
    \label{appen6}
\end{equation}
 Thus the total energy is automatically conserved with $Q_\phi$ and $Q_\chi$ as determined by 
 the Lagrangian equations without making an additional assumption. 
   Further, we note that 
 $Q_\phi+Q_\chi =\dot V_{12}$ and is definitely non-vanishing which would not be the case if we
 were to set  $Q_\phi=-Q_\chi=Q$, i.e., if one assumes that the sources are $\pm Q$ as is done is many works in the literature. Thus the $\pm Q$ assumption is inconsistent with field theory.  Next we exhibit the fluid equations in conformal time using  $\text{d}t =a\,\text{d}\tau$
 where $t$ is cosmic time and $\tau$ the conformal time. Using conformal time, the fluid
 equations now read
 \begin{align}
\label{appen3a}
    {\rho'}_\phi + 3{\cal H}(\rho_\phi + p_\phi) &= {\chi'} \frac{\partial V_{12}}{\partial \chi} 
    \equiv Q_\phi\,, \\
    {\rho'}_\chi + 3{\cal H}(\rho_\chi + p_\chi) &= {\phi'} \frac{\partial V_{12}}{\partial \phi}
    \equiv Q_\chi.
    \label{appen4a}
\end{align}
where prime means derivative with respect to conformal time and ${\cal H}= a H$.
Note that 
\begin{align}
Q_\phi+Q_\chi&=  {\chi'} \frac{\partial V_{12}}{\partial \chi} +  {\phi'} \frac{\partial V_{12}}{\partial \phi}= {V}'_3\,,
\label{appen5a}
\end{align}
and
\begin{equation}
    {\rho'} + 3{\cal H}(\rho + p) = 0.
    \label{appen6a}
\end{equation}
The analysis above 
 establishes the fact that the correct fluid equations in a field theoretic formalism for the case of
 two interacting fluids are given by  Eqs.~(\ref{appen3a}) and~(\ref{appen4a})
 where $Q_\phi+ Q_\chi\neq 0$ except for the case of no interaction between the fields.

\section{Background equations for a general dark matter potential}

We give the background evolution equations of dark matter and dark energy for the case of a general dark matter potential. For the interacting quintessence-dark matter model (QCDM), $\chi$ is a dark matter 
field while $\phi$ is a dark energy field. Instead of the DM potential of Eq.~(\ref{v1}), we consider the general form
\begin{align}
&V_1(\chi)=m_\chi^2 f^2\left[1+\cos\left(\frac{\chi}{f}\right)\right], 
\label{v1s}
\end{align}
while that potentials $V_2(\phi)$ and $V_{12}(\phi,\chi)$ have the forms of Eqs.~(\ref{v2}) and~(\ref{v3}). Define the DM energy density so that $\tilde{\rho}_\chi=\rho_\chi-V_{12}$ with the modified energy density fraction being $\tilde{\Omega}_\chi\equiv\tilde{\rho}_\chi/\rho_{\rm cr}$. We recast the KG equation for $\chi$ and $\phi$, Eqs.~(\ref{kgc0}) and~(\ref{kgp0}), in terms of the new variables defined by Eqs.~(\ref{var-tilde-1})$-$(\ref{var-tilde-3}). We then obtain the differential equations of $\Omega_\chi=\rho_\chi/\rho_{\rm cr}$, $\theta$ and $y$. For $\Omega_\chi$ we have 
\begin{align}
    \Omega^\prime_\chi&=3\mathcal{H}\Omega_\chi(w_T-w_\chi) 
    +\frac{\kappa^2 a^2}{3\mathcal{H}^2}\Bigg[\mathcal{H}(1+3w_\chi)V_{12}-\chi^\prime V_{12,\chi}\Bigg],
    \label{omc}
\end{align}
with the DM equation of state $w_\chi=-\cos\theta$ and $w_T=\sum p_i/\sum\rho_i$, where the sum is over all species (baryons, photons, neutrinos, DM and DE). As for the variables $\theta$ and $y$, we get
\begin{align}
    \theta^\prime&=-3\mathcal{H}\sin\theta+\mathcal{H}y 
    -\frac{\kappa^2 a^2}{3\mathcal{H}^2\tilde{\Omega}_\chi}\Bigg(2\mathcal{H} V_{12}+\chi^\prime V_{12,\chi}\Bigg)\cot\frac{\theta}{2}, 
    \label{thetac}
\end{align}
and
\begin{align}
    y^\prime&=\frac{3}{2}{\cal H}(1+w_T)y+\frac{\beta_\chi}{2}\mathcal{H}\tilde{\Omega}_\chi \sin\theta,
    \label{yeq}
\end{align}
where $\beta_\chi\equiv 3/(\kappa^2 f^2$) and $\tilde{\Omega}_\chi=\Omega_\chi-(\kappa^2 a^2/3\mathcal{H}^2)V_{12}$.
Note that the last set of terms  in each of Eqs.~(\ref{omc}) and~(\ref{thetac}) represent the interaction term. 

Next, we consider the assumption $Q_\chi=-Q_\phi$ and discuss its impact on cosmology from a field theory perspective. This assumption directly leads to $V_{12}^\prime=0$ so that 
\begin{equation}
    {V}_{12,\phi}\phi_0^\prime+{V}_{12,\chi}\chi_0^\prime=0,
    \label{v3p}
\end{equation}
Differentiating Eq.~(\ref{v3p}) with respect to time and using Eq.~(\ref{kgp0}) we get
\begin{align}
    &V_{12,\chi\chi}\chi_0^{\prime 2}+V_{12,\chi}\chi_0^{\prime\prime}+V_{12,\phi\phi}\left(\frac{V_{12,\chi}}{V_{12,\phi}}\right)^2\chi_0^{\prime 2}-2\mathcal{H}V_{12,\phi}\left(-\frac{V_{12,\chi}}{V_{12,\phi}}\chi_0^\prime\right)-a^2V_{12,\phi}V_{t,\phi}=0,
    \label{v3pp}
\end{align}
where $V_{t}=V_2+V_{12}$.
Now let us take Eq.~(\ref{v3p}) and differentiate one time with respect to $\chi$
\begin{equation}
    V_{12,\chi\chi}\chi_0^\prime+V_{12,\chi\phi}\phi_0^\prime=0,
\end{equation}
and then another with respect to $\phi$
\begin{equation}
    V_{12,\phi\chi}\chi_0^\prime+V_{12,\phi\phi}\phi_0^\prime=0.
\end{equation}
Eliminating $V_{12,\phi\chi}$ from both equations gives
\begin{equation}
    V_{12,\chi\chi}=V_{12,\phi\phi}\left(\frac{V_{12,\chi}}{V_{12,\phi}}\right)^2.
\end{equation}
Using this result in Eq.~(\ref{v3pp}), we get 
\begin{equation}
    \chi_0^{\prime\prime}+2\mathcal{H}\chi_0^\prime+\frac{2V_{12,\chi\chi}}{V_{12,\chi}}\chi_0^{\prime 2}-a^2\frac{V_{12,\phi}V_{t,\phi}}{V_{12,\chi}}=0,
    \label{kgc-new}
\end{equation}
which is a different equation from the KG equation of $\chi$, i.e., Eq.~(\ref{kgc0}). However, the field $\chi$ cannot obey two different KG equations. Setting Eq.~(\ref{kgc-new}) equal to Eq.~(\ref{kgc0}) gives
\begin{equation}
    \chi^{\prime 2}_0=\frac{a^2 V_{12,\chi}}{2V_{12,\chi\chi}}\left(V_{1,\chi}+ V_{12,\chi}+\frac{V_{12,\phi}V_{t,\phi}}{V_{12,\chi}}\right).
    \label{kin}
\end{equation}
Since $V_{12}=\frac{\lambda}{2}\chi^2\phi^2=\text{constant}$ implies  that 
{$\chi^2\phi^2=c$}, where $c$ is a constant of time. In this case, Eq.~(\ref{kin}) becomes
\begin{equation}
    \chi^{\prime 2}_0=\frac{a^2}{2}\chi_0\left(V_{1,\chi}+V_{12,\chi}+\frac{\chi_0}{\phi_0}V_{t,\phi}\right).
    \label{cpnew}
\end{equation}
Using Eq.~(\ref{om-tilde}), Eq.~(\ref{cpnew}) and Eqs.~(\ref{var-tilde-1})$-$(\ref{var-tilde-3}), we get a  new set of differential equations
\begin{align}
    \Omega_\chi^\prime&=3\mathcal{H}\tilde{\Omega}_\chi(1+w_T)-\frac{5}{8}\mathcal{H}\tilde{\Omega}_\chi y\sin\theta+\frac{\kappa^2 a^2}{2\cal H}\lambda c(1+w_T) \nonumber \\
    &+\frac{\sqrt{6}\kappa\,a^2}{12\mathcal{H}}\tilde{\Omega}_\chi^{1/2}\mathcal{V}_1\sin\frac{\theta}{2}-\frac{\kappa^2 a^2}{12\mathcal{H}^2}c\frac{\phi^\prime_0}{\phi^3_0}\mathcal{V}_2-\frac{\kappa^2}{48}c\, Y,    
    \label{omc-con}
\end{align}
and
\begin{align}
    \theta^\prime&=\frac{1}{8}\mathcal{H}y(3-5\cos\theta)+\frac{\sqrt{6}\kappa\,a^2}{12\mathcal{H}\tilde{\Omega}_\chi^{1/2}}\mathcal{V}_1\cos\frac{\theta}{2}-\frac{\kappa^2 a^2}{12\mathcal{H}^2\tilde{\Omega}_\chi}c\frac{\phi^\prime_0}{\phi^3_0}\mathcal{V}_2\cot\frac{\theta}{2}-\frac{\kappa^2}{48\tilde{\Omega}_\chi}c\,Y\cot\frac{\theta}{2},
    \label{theta-con}
\end{align}
where 
\begin{align}
\mathcal{V}_1&=\frac{2V_{2,\phi}}{\phi_0}+\frac{2\lambda\,c}{\phi^2_0}-V_{2,\phi\phi}, \\
\mathcal{V}_2&=\lambda\phi^2_0+\frac{V_{2,\phi}}{\phi_0}+\frac{\lambda\,c}{\phi^2_0}, \\
Y&=\frac{\phi^\prime_0}{\phi^3_0}\left[y^2-\beta_\chi \tilde{\Omega}_\chi (1+\cos\theta)\right].
\end{align}
Here one finds that the new Eqs.~(\ref{omc-con}) and~(\ref{theta-con}) are driven by the DE potential and the solution to the KG equation of $\phi$. Due to the constraint $V_{12}=\lambda c/2$, one can determine an equation for $y$ so that
\begin{align}
    y&=\frac{4}{\cal H}\frac{\phi^\prime_0}{\phi_0}\tan\frac{\theta}{2}-\frac{2\kappa^2 a^2}{3\mathcal{H}^3\tilde{\Omega}_\chi \sin\theta}c\frac{\phi^\prime_0}{\phi_0^3}\mathcal{V}_2.
    \label{yeq-new}
\end{align}

\section{Comparison between QCDM and the phenomenological model}

\begin{figure}[H]
\centering
\includegraphics[width=0.49\linewidth]{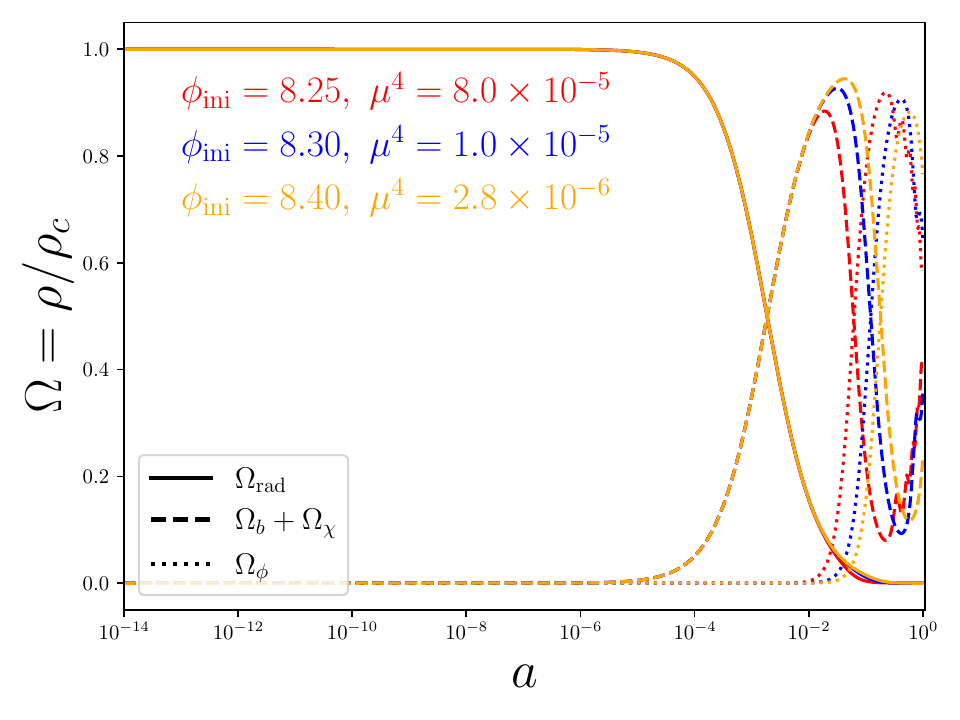}
\includegraphics[width=0.49\linewidth]{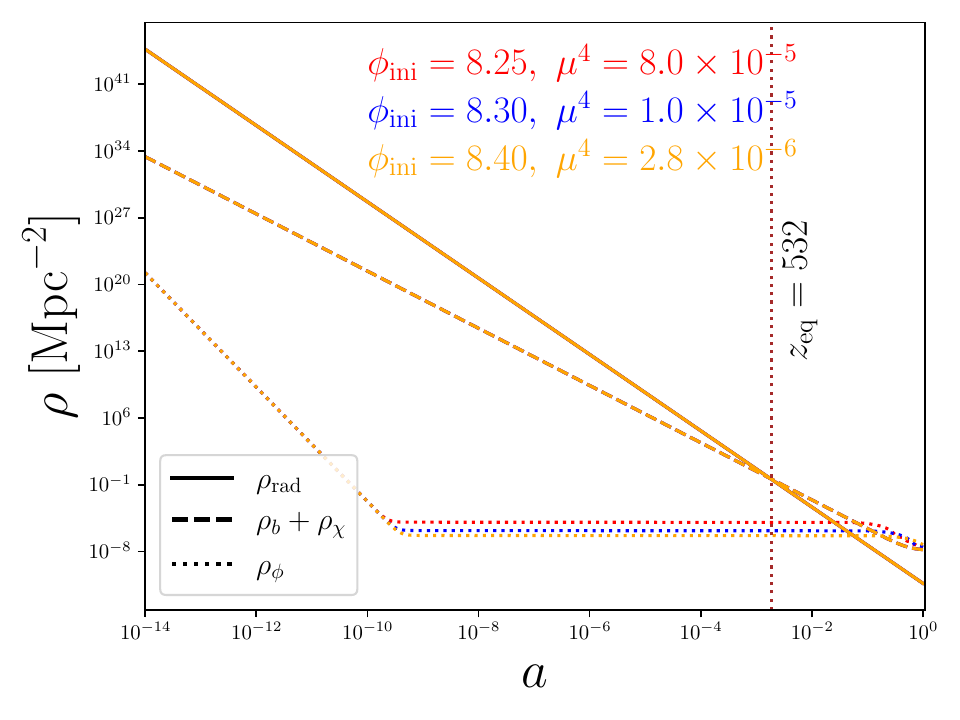}\\
\includegraphics[width=0.49\linewidth]{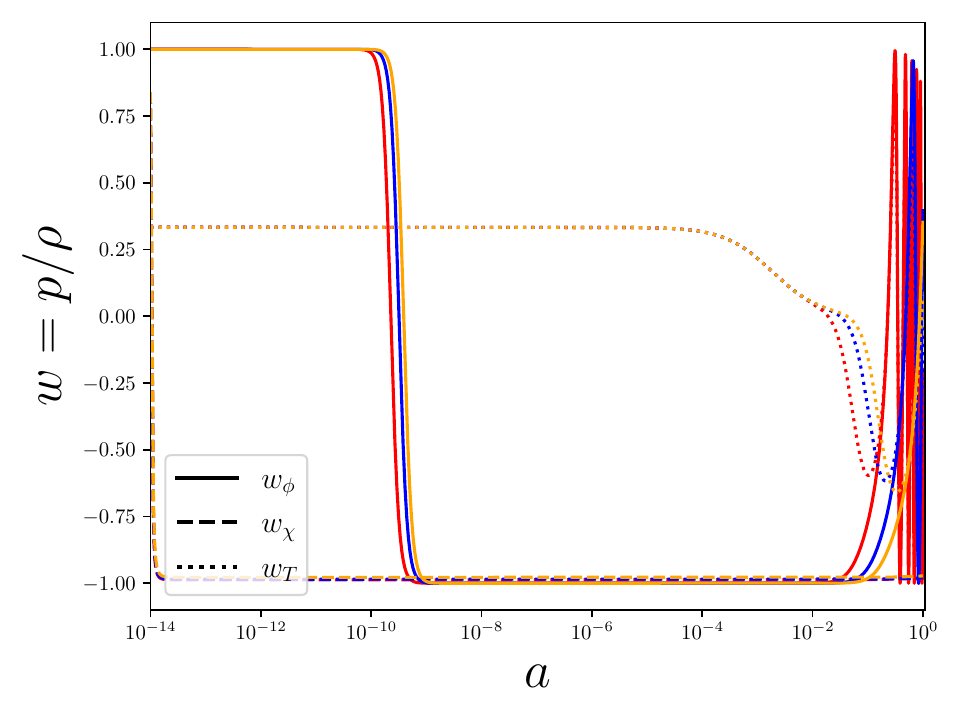}
\includegraphics[width=0.49\linewidth]{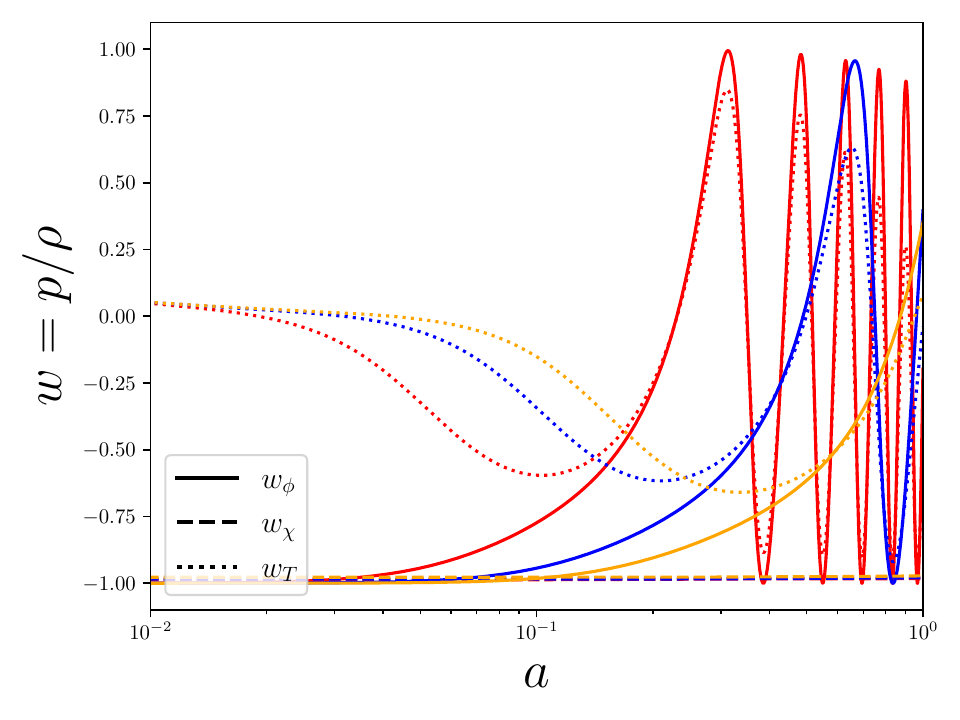}
\caption{\normalsize Results from the second scenario. Plots of the energy density fraction (top left) and the energy density (top right) of all species and the equations of state (EoS) of DM and DE and the total EoS (bottom panels) as a function of the scale factor $a$. The color code in the top panels is the same for all panels and correspond to different values of $\phi_{\rm ini}$ and $\mu^4$, expressed in standard \code{CLASS} units of $m_{\rm pl}$ and $m_{\rm pl}^2/\mathrm{Mpc}^2$, respectively.}
\label{fig1S}
\end{figure}

The phenomenological model based fluid equations assume that $Q_\chi=-Q_\phi$ which, as was shown in this work, is in contradiction with the field theory approach. To compare the results from this phenomenological model with that of QCDM, we show in Fig.~\ref{fig1S} the same plots of energy density and equations of state as in Fig.~\ref{fig1}. The obtained trend is similar to that of normal cosmology but here one can see an important difference relative to the QCDM analysis concerning matter-radiation equality which is attained at $z_{\rm eq}\sim 532$, in disagreement with Planck's result. Another important difference can also be seen from the evolution of the DM equation of state in the lower left panel of Fig.~\ref{fig1S} (dashed line). One can clearly see that the EoS takes a value of $-1$ throughout the evolution and so the DM field acts as DE. This is the main reason for getting an inconsistent cosmology resulting in a very late matter-radiation equality.

\section{Correlation plots}

\begin{figure}[H]
\begin{centering}
\includegraphics[width=0.8\textwidth]{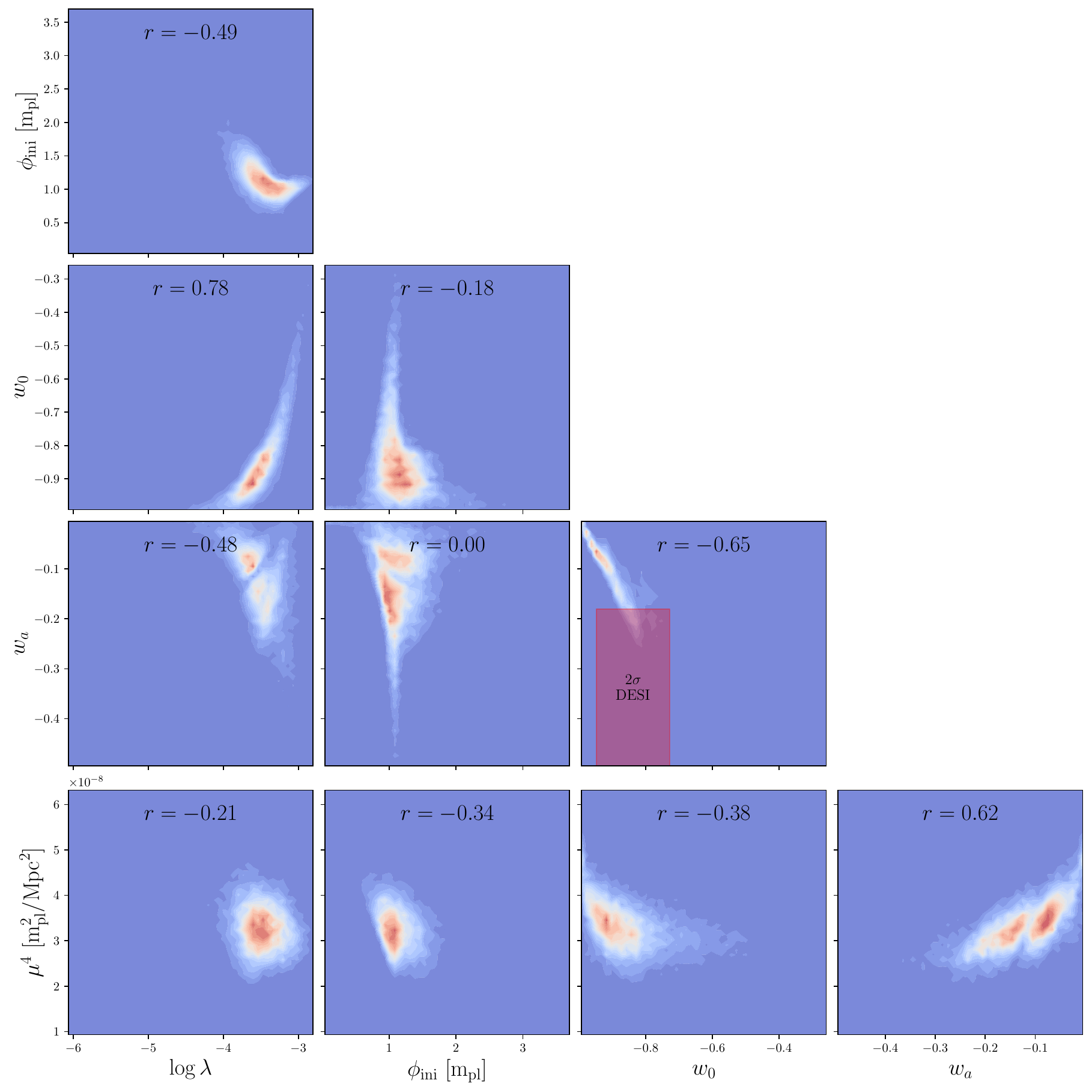}
\caption{Triangle plots showing the correlations between different input and output parameters of the model. Notice the strong positive correlation between $\lambda$ and $w_0$ and between $\mu^4$ and $w_a$. A mild negative correlation exists between $\lambda$ and $w_a$. }
\label{fig2S}
\end{centering}
\end{figure}

\begin{figure}[H]
\begin{centering}
\includegraphics[width=1.1\textwidth]{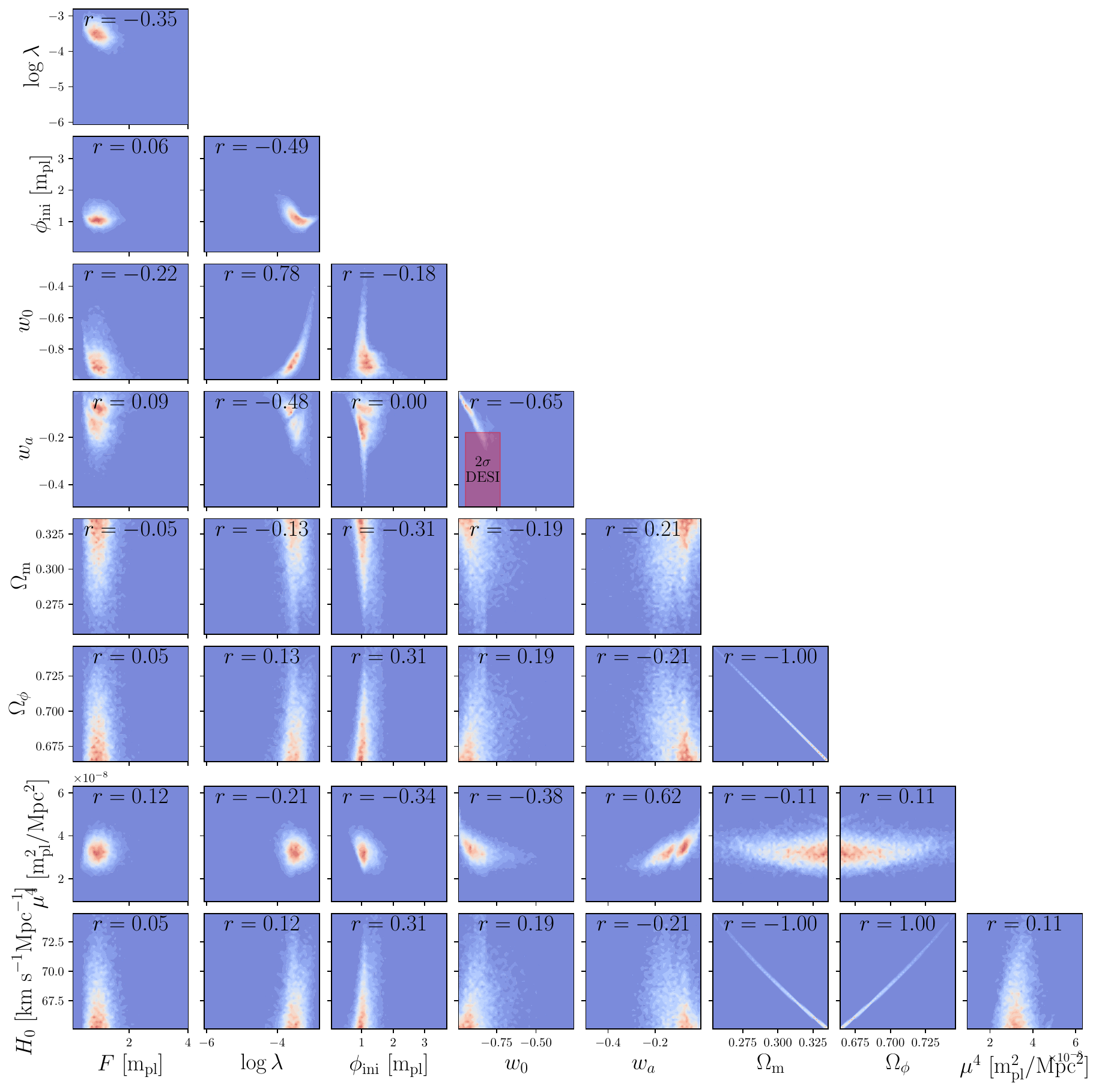}
\caption{Same as in Fig.~\ref{fig2S} but now considering a larger set of input and output parameters. }
\label{fig3S}
\end{centering}
\end{figure}

In Figs.~\ref{fig2S} and~\ref{fig3S}, we show the correlation plots between the different parameters of the model. From Fig.~\ref{fig2S}, one can see a strong positive correlation between $\lambda$ and $w_0$ but a weaker negative correlation between $\lambda$ and $w_a$. However, the latter has a stronger positive correlation with $\mu^4$. So one sees that the interplay between different parameters of the model can bring concordance with DESI DR2. A larger set of parameters is shown in Fig.~\ref{fig3S}. One important observation that can be made is that $H_0$ and $w_0$ are not anti-correlated as is usually seen in uncoupled quintessence models~\cite{Banerjee:2020xcn,Lee:2022cyh} as well as in some coupled quintessence~\cite{Das:2005yj}. A field-theoretic interaction term results in a very mild positive correlation between $H_0$ and $w_0$. This is, however, still not sufficient to explain the Hubble tension. 

\begin{figure}[H]
\begin{centering}
\includegraphics[width=1.0\textwidth]{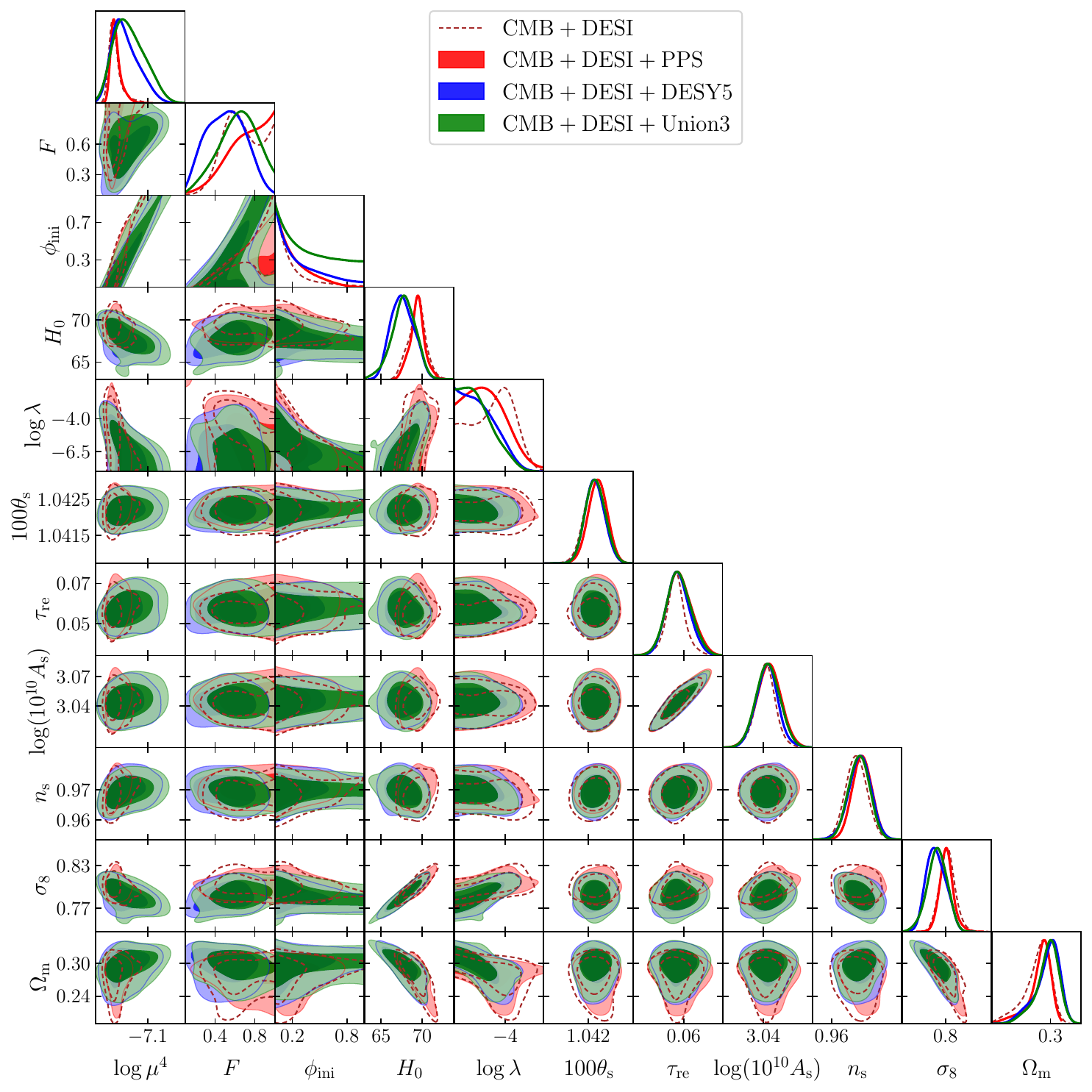}
\caption{Triangle plot showing the 2D joint and 1D marginalized posteriors of $\log\mu^4$, $\phi_{\rm ini}$, $\lambda$, $H_0$, $\tau_{\rm re}$, $\sigma_8$, $\Omega_{\rm m}$, $\theta_s$, $A_s$ and $n_s$ in the interacting dark matter-dark energy model for the following data set combinations: CMB+DESI (dashed brown contours), CMB+DESI+Pantheon+SH0ES (red contours), CMB+DESI+DESY5 (blue contours) and CMB+DESI+Union3 (green contours). }
\label{fig4S}
\end{centering}
\end{figure}

Fig.~\ref{fig4S} shows the different marginalized 2D posterior contours of the different cosmological parameters in QCDM using the full MCMC analysis with the inclusion of perturbations.

\end{document}